\documentclass[useAMS,usenatbib]{mn2e}

\usepackage{subfigure}
\usepackage{graphicx}
\usepackage{times}
\usepackage{tabularx}
\usepackage{natbib}
\usepackage{url} 
\usepackage{xcolor}
\usepackage{caption}
\usepackage{amsmath}
\usepackage{amssymb}
\usepackage{float}
\usepackage{multirow}
\usepackage{morefloats}

\newcommand{\enzo}{{\it {\small ENZO}}}

\newcommand{\music}{{\it {\small MUSIC}}}
\newcommand{\dd}{\mathrm{d}}
\newcommand{\Mpc}{\mathrm{Mpc}}
\newcommand{\Msun}{\mathrm{M}_{\odot}}

\newcommand{\kpc}{\mathrm{kpc}}
\newcommand{\radio}{\mathrm{radio}}

\newcommand{\km}{\mathrm{km}}
\newcommand{\sek}{\mathrm{s}}
\newcommand{\keV}{\mathrm{keV}}

\newcommand{\erg}{\mathrm{erg}}
\newcommand{\Hz}{\mathrm{Hz}}
\newcommand{\MHz}{\mathrm{MHz}}
\newcommand{\GHz}{\mathrm{GHz}}

\newcommand{\nuobs}{\nu_{\mathrm{obs}}}

\newcommand{\Cspec}{C_{\mathrm{spec}}}

\newcommand{\m}{\mathrm{m}}
\newcommand{\Jy}{\mathrm{Jy}}

\newcommand{\me}{m_\mathrm{e}}
\newcommand{\mpr}{m_\mathrm{p}}

\newcommand{\aint}{\alpha_{\mathrm{int}}}

\definecolor{myred}{rgb}{1,0,0} 
\definecolor{myblue}{rgb}{0,0,1}
\definecolor{mygreen}{rgb}{0,1,0}

\begin{document}

 \title[Spectral properties of radio relics]{Exploring the spectral properties of radio relics I: \\Integrated spectral index and Mach number}
 \author[D. Wittor et al.]{D. Wittor$^{1,2}$\thanks{%
 E-mail: dwittor@hs.uni-hamburg.de}, S. Ettori$^{3,4}$, F. Vazza$^{2,1,5}$, K. Rajpurohit$^{2,5,6}$, M. Hoeft$^{6}$,P. Dom\'{i}nguez-Fern\'{a}ndez$^{7,1}$  \\
 %EndANameFCm
 $^{1}$ Hamburger Sternwarte, Gojenbergsweg 112, 21029 Hamburg, Germany \\
 $^{2}$ Dipartimento di Fisica e Astronomia, Universita di Bologna, Via Gobetti 93/2, 40122, Bologna, Italy \\
 $^{3}$ INAF, Osservatorio di Astrofisica e Scienza dello Spazio, via Pietro Gobetti 93/3, 40129 Bologna, Italy \\
 $^{4}$ INFN, Sezione di Bologna, viale Berti Pichat 6/2, 40127 Bologna, Italy \\
 $^{5}$ INAF, Istituto di Radioastronomia di Bologna, via Gobetti 101, I-41029 Bologna, Italy \\
 $^{6}$ Th\"uringer Landessternwarte, Sternwarte 5, 07778 Tautenburg, Germany \\
 $^{7}$ Department of Physics, School of Natural Sciences UNIST, Ulsan 44919, Republic of Korea
  }
 \date{Accepted ???. Received ???; in original form ???}
 \maketitle

 \begin{abstract}
 Radio relics are the manifestation of electrons presumably being shock (re-)accelerated to high energies in the outskirts of galaxy clusters. However, estimates of the shocks' strength yield different results when measured with radio or X-ray observations. In general, Mach numbers obtained from radio observations are larger than the corresponding X-ray measurements. In this work, we investigate this \textit{Mach number discrepancy}. 
 For this purpose, we used the cosmological code \enzo \ to simulate a sample of galaxy clusters that host bright radio relics. For each relic, we computed the radio Mach number from the integrated radio spectrum and the X-ray Mach number from the X-ray surface brightness and temperature jumps. Our analysis suggests that the differences in the Mach number estimates follow from the way in which different observables are related to different parts of the underlying Mach number distribution: radio observations are more sensistive to the high Mach numbers present only in a small fraction of a shock's surface, while X-ray measurements reflect the average of the Mach number distribution.  Moreover, X-ray measurements are very sensitive to the relic's orientation. If the same relic is observed from different sides, the measured X-ray Mach number varies significantly. On the other hand, the radio measurements are more robust, as they are unaffected by the relic's orientation.
 \end{abstract}
 \label{firstpage}
 
 \section{Introduction}
 
 Radio relics and halos, diffuse radio emission observed in galaxy clusters, illuminate shock fronts and turbulence caused by cluster mergers \citep[e.g.][]{2008SSRv..134...93F,vanweeren2019review}. While releasing energies of about $\gtrsim 10^{64}  \ \erg$, cluster mergers induce shock waves and turbulence in the in\-tra\-clus\-ter medium (ICM), the hot ionised plasma that resides between the galaxies. Radio halos are believed to be produced by turbulence \citep[e.g.][]{2014IJMPD..2330007B,2015A&A...580A..97C}. On the other hand, radio relics are are usually located at the clusters' periphery where they are often found to be associated with discontinuities in the X-ray surface brightness (or more rarely, temperature jumps) plausibly connected with shock waves \citep[e.g.][]{2002ApJ...567L..27M,2016MNRAS.463.1534B,2017A&A...600A.100A}. Such observations supported the idea that the shock waves accelerate thermal electrons of the ICM to high energies via diffusive shock acceleration (DSA) \citep[e.g.][]{1983RPPh...46..973D,ensslin1998,2013ApJ...764...95K}. The cosmic-ray electrons, that are embedded in the large-scale magnetic fields, emit synchrotron radiation, forming radio relics \citep[e.g.][]{2014IJMPD..2330007B,bykov2019review}.
 
 While the shock acceleration scenario is widely accepted as the origin of relics, the viability of the DSA mechanism is still questioned. Since many years, the non-detection of the $\gamma$-ray signal from accelerated cosmic-ray protons is posing a problem to the model \citep[e.g.][]{va14relics,Ryu2019,Ha2019gammas,wittor2020gammas,wittor2020review}.  Still, the acceleration efficiencies for DSA from the thermal pool are too low to explain the observed high radio powers of most relics \citep[e.g.][]{2020A&A...634A..64B}, in particular if the Mach number of the shock is as low as derived from X-ray.
 
 Using X-ray observations, shocks are measured as jumps in the surface brightness or temperature profiles. The Mach number of the underlying shock is computed from the deprojection of the measured profiles  \citep[e.g.][]{Markevitch:2007}. In general, the obtained Mach numbers are fairly low with values of $M_{\mathrm{X-ray}} \approx 1.5 - 2.5$ \citep[e.g.][and references in Table \ref{tab::observations}]{2011MmSAI..82..495M,2013PASJ...65...16A,2016MNRAS.461.1302E}. 
 
 Radio observations allow to derive the shock strength from the integrated radio spectrum. Under the assumption of stationary shock conditions in the ICM, the radio spectral index depends on the shock's Mach number. In most cases, the observed radio Mach numbers yield larger values than the corresponding X-ray Mach number with $M_{\radio} \approx 2 - 5$ \citep[e.g.][]{2016ApJ...818..204V,2018MNRAS.478.2218H,Stuardi2019}.
 
 This tension may impose a challenge on the DSA scenario for radio relics. Several works have tried to explain this Mach number discrepancy, which could be inherent to biases in the observations as well as to properties of the ICM or the underlying acceleration mechanism. As seen both in simulations and observations, relics are not planar and uniform structures, but their morphologies are complex and show non-linear features. As a consequence, the Mach number must be carefully determined from the observed 2D maps. \citet{2017A&A...600A.100A} discussed various systematic errors that could bias the X-ray observations. Biases in the radio observations have been discussed in \citet{2016ApJ...818..204V} and \citet{Hoang2017sausage}. We discuss these observational biases in Section \ref{ssec::obs}. Furthermore, modifications to the basic DSA theory have been proposed to solve the Mach number discrepancy. These modifications include the re-acceleration of a population of mildly relativistic electrons \citep[e.g.][]{2005ApJ...627..733M,2015ApJ...809..186K,2016MNRAS.455.2402S}, Alfv\'enic drifts \citep{2018ApJ...856...33K}, $\kappa$-distributions for the suprathermal electrons \citep{2014ApJ...788..142K} or superdiffusion \citep{Zimbardo2018}. 
   
 In this work, we study the Mach number discrepancy found in radio relics using synthetic shock fronts as well as 3D cosmological simulations. Specifically, we focus on the underlying Mach number distribution and how it is reflected in the radio and X-ray Mach number. Various cosmological simulations have shown that a radio relic does not trace a single Mach number but a distribution of Mach numbers \citep[e.g.][]{2011JApA...32..509H,2013ApJ...765...21S,wittor2019pol,Roh2019}. \citet{Dominguez_2020_relicsI} showed that such a distribution is produced if a shock wave with an initially uniform Mach number runs into a turbulent medium with fluctuations as found in the ICM \citep[e.g.][]{2011ApJ...731L..10N,2011Sci...331.1576S,2021arXiv210201096A}. 
 Here, we tackle the following questions: \textit{How does the distribution of Mach numbers in the shock front and the viewing angle affect the Mach number estimates from X-ray and radio observations? Does this explain the Mach number discrepancy?}
  
 We structured our work as followed: in the next section, we present an overview of the available observations. In Section \ref{sec::equations}, we present the equations that we used to model the radio emission produced by shock waves. Furthermore, we introduce a variety of quantities that we used for our analysis. In Section \ref{sec::1D}, we build synthetic shock fronts to study the Mach number discrepancy in a controlled environment. In Section \ref{sec::3D}, we study the Mach number discrenpancy in cosmological simulations. We summarise and discuss our work in Section \ref{sec::summary}. The conclusion is given in Section \ref{sec::conclusion}. We note that throughout this work, we refer to the Mach number measured in X-ray as X-ray Mach number and to the one measured in radio as radio Mach number.

 \section{Mach Number Discrepancy of Radio Relics} \label{ssec::obs}
 
 For a significant number of observed radio relics, deep enough X-ray and radio data exists to estimate the Mach number of the underlying shock wave in both wavelengths. By the time of writing, 29 relics and their corresponding shock waves have been detected in radio and X-ray, respectively. Out of these 29 relics, we compiled a list of 21 relics that have a clear evidence for a shock wave in X-rays and that is located at the radio relic. In Table \ref{tab::observations}, we summarise the main properties of the selected relics. 
 
 \begin{table*}
 \begin{tabular}{lcccccccc}
 Name & $z$ & $S_{1.4 \ \GHz}$ & $\aint$ & $M_{\mathrm{radio}}$ & $M_{\mathrm{X-ray}}$ & $M_{\mathrm{X-ray}}$ & radio ref. & X-ray ref. \\ 
 & & $[\mathrm{mJy}]$ &  & $(\aint)$ & $(\Delta T)$ & $(\Delta \rho)$ &  &  \\ 
\hline
RXCJ1314.4-2515 W & 0.247 & $20.2 \pm 0.5$ & $1.22^{+0.09}_{-0.09}$ & $3.18^{+0.59\#}_{-0.59}$ &  $2.4^{+1.10}_{-0.80}$ & $1.7^{+0.40}_{-0.20}$ & rA & xA \\
  &  &  &  &  & n/a & $2.1^{+0.10}_{-0.10}$ &  & xB \\
Abell 521 & 0.247 & $14.0^{a} \pm 1.0 $ & $1.48^{+0.01}_{-0.01}$ & $2.27^{+0.02}_{-0.02}$ & $3.4^{+3.69}_{-1.69}$ & $2.42^{+0.19}_{-0.19}$ & rB & xC \\
Abell 2146 & 0.232 & $1.1^{b}$ & $1.14^{+0.08}_{-0.08}$ & $3.91^{+1.04\#}_{-1.04}$ & $2.0^{+0.30}_{-0.30}$ & $1.6^{+0.10}_{-0.10}$ & rC & xD \\
El Gordo NW & 0.87 & $8.8^* \pm 0.9$ & $1.37^{+0.20}_{-0.20}$ & $2.53^{+1.04,c}_{-0.41}$ & $<2.9^{d}$ &  $2.78^{+0.63,e}_{-0.38}$ & rD &  xE \\
 &  &  &  & & n/a & $2.4^{+1.30,f}_{-0.60}$ & & xF \\
ZwCL 2341.1+0000 NW  & 0.27 & 5.0 & 1.2 & 3.3 & n/a & $2.06^{+1.39}_{-0.76}$ & rE & xG \\
ZwCL 2341.1+0000 SE  & 0.27 & 13.0 & 1.2 & 3.3 & n/a & $1.43^{+0.23}_{-0.20}$ & rE & xG \\
Bullet Reg. B & 0.296 & $4.8^{g} \pm 0.6$ & $1.70^{+0.35}_{-0.31}$ & $1.96^{+0.32}_{-0.36}$ & $2.5^{+1.30}_{-0.80}$ & n/a & rF & xH \\
Toothbrush & 0.225 & $335.8^{h} \pm 22.8$ &  $1.16^{+0.03}_{-0.03}$ & $3.7^{+0.30}_{-0.30}$  & $<1.5^{i}$ & $1.2^{+0.13,j}_{-0.13}$  & rG & xI \\
 &  & &  &   & n/a & $1.5^{+0.37}_{-0.27}$  &   & xJ \\
Sausage North & 0.192 & $126.0^{k} \pm 12.6$ &  $1.12^{+0.30}_{-0.30}$ & $4.2^{+0.40}_{-0.60}$ & $2.7^{+0.70}_{-0.40}$ & n/a & rH & xK \\
Sausage South & 0.192 & $18.0^* \pm 0.5$ & $1.12^{+0.07}_{-0.07}$ & $4.2^{+1.16\#}_{-1.16}$  & $1.7^{+0.40}_{-0.30}$ & n/a & rI & xK \\
Abell 2744 R1 & 0.308 & $13.3^* \pm 1.1$ & $1.17^{+0.03}_{-0.03}$ &  $3.6^{+0.30}_{-0.20}$ & n/a & $1.7^{+0.5}_{-0.3}$ & rJ & xL \\
      & &  &  & & $3.7^{+0.40}_{-0.40}$ & n/a &   & xM  \\
 Abell 2744 R2 & 0.308 & $2.1^* \pm 0.2$ & $1.19^{+0.05}_{-0.05}$ & $3.4^{+0.41\#}_{-0.41}$ & n/a & $1.26^{+0.25}_{-0.15}$ & rJ & xN \\
Abell 1240-1 & 0.159 & $5.5^* \pm 0.7$ & $1.08^{+0.05}_{-0.05}$ & $5.1^{+3.10}_{-1.10}$ & n/a & 2.0 & rK & xO \\
Abell 1240-2 & 0.159 & $13.3^* \pm 1.5$ & $1.13^{+0.05}_{-0.05}$ & $4.0^{+1.10}_{-0.60}$ & n/a & 2.0 & rK & xO \\
Abell 754 & 0.054 & $6.0 \pm 0.3$ & $1.77^{+0.10}_{-0.10}$ & $1.90^{+0.09\#}_{-0.09}$ & n/a & $1.57^{+0.16}_{-0.12}$ & rL  & xP  \\
                 & & &  &  &  n/a & $1.71^{+0.45}_{-0.28}$ &  & xQ \\
Abell 115 & 0.197 & 147.0 &  $1.1^{+0.50}_{-0.50}$ & $4.58^{+10.9\#}_{-10.9}$ & $1.7^{+0.10}_{-0.10}$ & $1.8^{+0.50}_{-0.40}$ & rM  & xR  \\
Abell 3376 West  & 0.046  & $166.0^{l}$ & $1.17^{+0.06}_{-0.06}$ & $3.57^{+0.58}_{-0.58}$ & $2.8^{+0.4}_{-0.4}$ & n/a & rN  & xS  \\
                 & & &   &   & $2.94^{+0.60}_{-0.60}$ & n/a &   & xT  \\
Abell 3376 East  & 0.046 & $136.0^{l}$ & $1.37^{+0.08}_{-0.08}$ & $2.53^{+0.23}_{-0.23}$ & $1.5^{+0.10}_{-0.10}$ & n/a & rN  & xS  \\
Coma & 0.023 & $240.5^{*} \pm 21.5$ &  $1.18^{+0.02}_{-0.02}$ & $3.4^{+0.18\#}_{-0.18}$ & $2.2^{+0.50}_{-0.20}$ & n/a & rO  & xT \\
     & & &   &  & $1.6^{+0.16}_{-0.40}$ &  n/a &  & xU  \\
 ZwCl 0008.8+5215 West & 0.104 & $10.8^{*} \pm 1.2$ & $1.59^{+0.06}_{-0.06}$ & $2.10^{+0.08\#}_{-0.08}$ & $2.35^{+0.74}_{-0.55}$ & $1.48^{+0.50}_{-0.32}$ & rP & xV \\
  & &  &  &  & $2.02^{+0.74}_{-0.43}$ & & & xV \\
 ZwCl 0008.8+5215 East & 0.104 & $54.9^{*} \pm 3.5$ & $1.49^{+0.12}_{-0.12}$ & $2.25^{+0.22\#}_{-0.22}$ & $1.54^{+0.65}_{-0.47}$ & n/a & rQ & xV \\
\end{tabular}
\caption{The table shows various properties of relics that have a Mach number measurement in both X-ray and radio. The columns provide: the cluster name, the redshift, the flux at $1.4 \ \GHz$, the integrated spectral index, the radio Mach number computed from the integrated spectral index, the X-ray Mach number estimated from the temperature and/or density jump, the references for the radio and X-ray Mach number. \\
We note that some works, did not provide the flux at $1.4 \ \GHz$. In these cases, we interpolated from other flux values, using the given integrated spectral index. The corresponding error was computed using error propagation. The fluxes that we interpolated are marked by the $*$. For Abell 2146, ZwCL 2341.1+0000, Abell 115 and Abell 3376 the flux values, we could not find the errors of the flux measurements. \\
Moreover, some works did not give the Mach number computed from the integrated radio spectrum. For these cases, we compute the radio Mach number using Eq. \ref{eq::ainj_mach} and the corresponding error via error propagation. These cases are marked with an $\#$. In the case of ZwCL 2341.1+0000, we did not find errors for neither the spectral index nor the Mach number. For Abell 1240, we did not find errors for the X-ray Mach number. \\
Finally, we have some more comments about specific works:
a: taken from \citet{2006NewA...11..437G};
b: taken from \citet{2018MNRAS.475.2743H};
c: errors taken from \citet{2020A&A...634A..64B};
d: taken the upper limit given in \citet{2016MNRAS.463.1534B};
e: X-ray surface brightness jump taken from \citet{2020A&A...634A..64B};
f: \citet{2016ApJ...829L..23B} measured the Mach number using the density jump and Sunayeav-Zeldovich measurements;
g: taken from \citet{2015MNRAS.449.1486S};
h: interpolated from the flux at $1.5 \ \GHz$ given in \citet{rajpurohit2020toothbrush};
i: taken the upper limit given in \citet{2016ApJ...818..204V};
j: error computed from the values given in section 4.3 of \citet{2016ApJ...818..204V} using error propagation
k: taken from \citet{2017MNRAS.472.3605L};
l: taken from \citet{2006Sci...314..791B};
\\
References:
rA: \protect{\citet{2017MNRAS.467..936G}};
rB: \protect{\citet{2008A&A...486..347G}};
rC: \protect{\citet{Hoang_2019_A2146}};
rD: \protect{\citet{2016MNRAS.463.1534B}};
rE: \protect{\citet{2010A&A...511L...5G}};
rF: \protect{\citet{2016Ap&SS.361..255M}};
rG: \protect{\citet{rajpurohit2020toothspec}};
rH: \protect{\citet{loi2020sausspec}};
rI: \protect{\citet{digennaro2018saus}};
rJ: \protect{\citet{rajpurohit2021A2744}};
rK: \protect{\citet{2018MNRAS.478.2218H}};
rL: \protect{\citet{2011ApJ...728...82M}};
rM: \protect{\citet{2001A&A...376..803G}};
rN: \protect{\citet{2015MNRAS.451.4207G}};
rO: \protect{\citet{2003A&A...397...53T}};
rP: \protect{\citet{2011A&A...528A..38V}};
rQ: \protect{\citet{2017A&A...600A..18K}}; \\
xA: \protect{\citet{Stuardi2019}};
xB: \protect{\citet{2011MmSAI..82..495M}};
xC: \protect{\citet{2013ApJ...764...82B}};
xD: \protect{\citet{2011MNRAS.417L...1R}};
xE: \protect{\citet{2016MNRAS.463.1534B}};
xF: \protect{\citet{2016ApJ...829L..23B}};
xG: \protect{\citet{2014MNRAS.443.2463O}};
xH: \protect{\citet{2015MNRAS.449.1486S}};
xI: \protect{\citet{2016ApJ...818..204V}};
xJ: \protect{\citet{2015PASJ...67..113I}};
xK: \protect{\citet{2015A&A...582A..87A}};
xL: \protect{\citet{2016MNRAS.461.1302E}};
xM: \protect{\citet{2017PASJ...69...39H}};
xN: \protect{\citet{2017ApJ...845...81P}};
xO: \protect{\citet{2018MNRAS.478.2218H}};
xP: \protect{\citet{2011ApJ...728...82M}};
xQ: \protect{\citet{2003AstL...29..425K}};
xR: \protect{\citet{2016MNRAS.460L..84B}};
xS: \protect{\citet{2018A&A...618A..74U}};
xT: \protect{\citet{2013PASJ...65...16A}};
xU: \protect{\citet{2013MNRAS.433.1701O}};
xV: \protect{\citet{2019ApJ...873...64D}}
}
  \label{tab::observations}
 \end{table*}
 
 In the compiled sample, the clusters range in redshift from $z = 0.023$ to $z = 0.87$ \citep[e.g.][]{2003A&A...397...53T,2012ApJ...748....7M}. At $1.4 \, \GHz$, the brightest relic of the sample is the Toothbrush, $S_{1.4 \, \GHz} = (335.8 \pm 22.8)  \, \m\Jy$ \citep{rajpurohit2020toothbrush}. With a flux of $S_{1.4 \, \GHz} = (1.1 \pm 0.5 )\, \m\Jy$, the relic in Abell 2146 is the faintest one of the sample \citep{2018MNRAS.475.2743H}. Abell 1240 has the flattest integrated spectral index of the sample, i.e. $\aint = 1.08 \pm 0.05$, while the one of Abell 754 is the steepest, i.e. $\aint = 1.77 \pm 0.1$ \citep[][respectively]{2018MNRAS.478.2218H,2011ApJ...728...82M}. 
 
 In DSA, particles are accelerated by crossing a shock front multiple times. This results in a power-law energy distribution of relativistic electrons, $n_E \propto E^{-s}$, whose slope, $s$, is only related to the shock's strength \citep[e.g.][]{1983RPPh...46..973D}:
\begin{align}
  s = 2 \frac{M^2+1}{M^2-1} \label{eq::s_mach}
\end{align}
 Following the theory of synchrotron emission, this population of electrons produces radio emission that is also described by a power-law (with an exponential cut-off at high frequencies, giving the high energy cut-off of the particle distribution). 
 In the standard model of radio relics \citep[e.g.][]{ensslin1998,2007MNRAS.375...77H}, the radio Mach number can be estimated from the integrated radio spectrum that follows a power-law that solely depends on the Mach number:
  \begin{align}
   I_{\nuobs} \propto \nuobs^{- \aint}, \ \ \mathrm{with} \ \ \aint = \frac{M^2 + 1 }{M^2 - 1 } \label{eq::ainj_mach},
 \end{align} 
 where $\aint$ is the spectral index at observing frequency $\nuobs$ and flux density $I_{\nuobs}$. 
 
 However, deriving Eq. \ref{eq::ainj_mach} from Eq. \ref{eq::s_mach} is not universal, but it depends on specific assumptions about the properties of the shock and the observing frequencies. The shock has to planar and both the magnetic field and the velocity in the downstream are assumed to be uniform. Furthermore at the observing frequencies, the downstream width is determined by radiative cooling. Furthermore, the model assumes that all frequencies of the observation are well below the critical synchrotron frequency of the maximum energy electrons injected by DSA. Finally, all electrons only experience radiative energy losses. With these assumptions, the integrated spectral indices of the sample translate into radio Mach numbers ranging from $M \approx 4.58$ to $M \approx 1.86$. However, if any of these assumptions are violated, other outcomes than Eq. \ref{eq::ainj_mach} are possible. For example, it has been shown that the slope of the radio spectrum exhibits non-linear features for non-planar shocks, such as spherical expanding ones \citep[][]{2015ApJ...809..186K,2015JKAS...48..155K,Dominguez_2020_relicsI}. However, the majority of observed relics follow the stationary shock condition and the Mach numbers obtained by radio colour-colour plots are consistent with those obtained from the integrated radio spectra \citep[e.g.][]{digennaro2018saus,rajpurohit2020toothbrush,rajpurohit2021macsspec}.
 
 Moreover, the integrated spectrum and the radio Mach number can be difficult to measure. Only for the minority of relics, a radio spectrum is measured at multiple frequencies \citep[e.g.][]{loi2020sausspec,rajpurohit2020toothspec,2020A&A...642A..85D}. Several radio measurements are based on three flux measurements only \citep[e.g.][]{2011ApJ...728...82M}. On top of that, if the radio observations are too shallow, they might miss some flux and, hence, underestimate the Mach number obtained from the overall spectrum. Yet, for most observations, the errors on the integrated spectral index are small.
 
 X-ray Mach numbers can be computed from either the deprojected temperature or density jump \citep{Markevitch:2007}. However, we did not find both X-ray Mach numbers for all relics. If both X-ray Mach numbers are available, the temperature based measurements give larger values, possibly indicating a systematic effect.
 Difficulties in the X-ray measurements can arise during the determination of the pre- and post-shock temperature \citep[e.g.][and references therein]{2017A&A...600A.100A}. If a relic is seen at an oblique viewing angle, the pre-shock gas temperature can be over-estimated. Consequently, both the observed temperature jump and the measured X-ray Mach number would be under-estimated. Cosmological simulations by \citet{2013ApJ...765...21S} and \citet{2015ApJ...812...49H} confirmed this effect. A similar effect occurs if X-ray observations under-estimate the post-shock temperature. For example, if the electrons do not reach thermal equilibrium, in case the equilibration time for electrons and protons is very different in a weakly collisional plasma  \citep{2017A&A...600A.100A}. 
  
 For about half of the relics, the X-ray observation measured different Mach numbers than the corresponding radio observation. In these cases, the X-ray Mach numbers mostly yield lower values than the radio Mach number. 
 
 In summary, observed relics present a wide range of both X-ray and radio Mach numbers. However, for about half of the relics, the two measurements do not match. The accurate measurement of Mach numbers from both techniques is prone to technical challenges that can bias the Mach number estimates. Such uncertainties could in principle explain the Mach number discrepancies in some relics. However, for a significant fraction of relics, the discrepancies are large and they should be of physical nature, which we investigate in the following. Throughout this work, we highlight the different properties of the observed sample in greater detail.
 
 We note that we did not include a few relics (namely Abell 520, Abell 2255, Abell 2256, Abell 3411 and Abell 3667) in our catalogue. In Abell 520, the relic overlaps with the radio halo \citep{2019A&A...622A..20H,2016ApJ...833...99W}, which makes the determination of the relic's integrated radio spectrum and corresponding radio Mach number more difficult. For the other four clusters, the assumption of quasi-stationary shock conditions is questionable. Abell 3411 shows a clear connection to a nearby AGN and the integrated spectrum is produced by a population of fossil electrons \citep{2017NatAs...1E...5V,2019ApJ...887...31A,2020A&A...642A..89Z}. In Abell 3667, Abell 2255 and Abell 2256, the measured integrated spectral indices of the radio relics are below $\leq 1$, already being in tension with DSA from the thermal pool, if the quasi-stationary radio shock condition applies  \citep{2014MNRAS.445..330H,2013AN....334..346S,2009A&A...507..639P,2013PASJ...65...16A,2018MNRAS.479..553S,2015A&A...575A..45T}.

\section{Models and Methods}\label{sec::equations}
 
 The overall goal of this work to study the Mach number discrepancy in 3D cosmological simulations, that realistically reproduce shock fronts caused by cluster mergers. We analyse a set of such simulations by assigning radio emission to the shock front. In this section, we describe the assumptions and the equations that we used to model the radio emission. Furthermore, we introduce some statistical measures that we used throughout our analysis.

 \subsection{Radio Emission}

 To compute the radio emission from relics, we apply the model derived by \citet{2007MNRAS.375...77H}. In \citet{wittor2019pol}, we used this model to study the polarised emission of radio relics. In the following, we briefly summarise the main equations. For details, we point to the given references.
 
 We computed the aged electron spectrum at time $t$ after the injection following \citet[][]{Kardashev1962}
  \begin{align} 
   \begin{split}
      n_{\mathrm{E}}&(E, t) =  \frac{n_{\mathrm{e}} C_{\mathrm{spec}} }{\me c^2} \left(\frac{E}{\me c^2}\right)^{-s}   \\
      &\times \left[1-\left(\frac{\me c^2}{E_{\max}} + C_{\mathrm{cool}}t\right)\frac{E}{\me c^2} \right]^{s-2}.
   \end{split} \label{eq::ne}
  \end{align}
 Here, $s$ is the slope of the energy spectrum, see Eq. \ref{eq::s_mach}, and $ C_{\mathrm{cool}}$ is the cooling constant that accounts for synchrotron radiation and inverse Compton losses
 \begin{align}
     C_{\mathrm{cool}} = \frac{\sigma_{\mathrm{T}}}{6 \pi \me c} \left(B^2_{\mathrm{CMB}}+B^2\right).
 \end{align}
 In the equations above, $\me$ is the electron mass, $c$ is the speed of light, $B$ is the magnetic field strength, $B_{\mathrm{CMB}}$ is the equivalent magnetic field of the cosmic microwave background at redshift $z$, $\sigma_{\mathrm{T}}$ is the Thomson cross-section and $n_e$ is the number density of electrons. The model uses both a constant velocity and a constant magnetic field in the downstream region. $E_{\max}$ is the maximum energy to which particles can be accelerated and it is chosen such that the corresponding critical synchrotron frequency is above the observing frequencies used in this work. The normalisation of the spectrum is 
 \begin{align}
     \Cspec = \xi_{\mathrm{e}} \frac{u_{\mathrm{d}}}{c^2}\frac{\mpr}{\me}\frac{q-1}{q}\frac{1}{I_{\mathrm{spec}}}, \label{eq::cspec}
 \end{align}
 and it depends on the internal energy of the downstream gas $u_{\mathrm{d}}$, the proton mass $\mpr$ and the entropy jump $q$. $\xi_{\mathrm{e}}$ is the fraction of energy that goes into the acceleration of electrons to suprathermal energies. The integral $I_{\mathrm{spec}}$ is:
 \begin{align}
     I_{\mathrm{spec}} = \int_{E_{\min}}^{\infty} E \left(\frac{E}{\me c^2}\right)^{-s} \left(1-\frac{E}{E_{\max}}\right)^{s-2} \dd E.
 \end{align}
 Electrons are considered to be suprathermal and to undergo shock acceleration if their energy is above $E_{\min} = 10 \mathrm{k_B} T$ and they can be accelerated to the finite energy $E_{\max} = 10^{10} E_{\min}$. At distance $x$ from the shock front, the associated radio power is computed as the integral of the electron spectrum and the modified Bessel functions $F(1/\tau^2)$
 \begin{align}
     \frac{\dd P}{\dd V \dd \nu }(x) &= C_{\mathrm{R}}  \int_{0}^{E_{\max}} n_{\mathrm{E}}(\tau, x) F\left(\frac{1}{\tau^2}\right) \dd \tau, \label{eq::dPdVdv} 
 \end{align}
where we substituted $t = x / v_d$ in the computation of the spectrum, with $v_d$ being the downstream velocity. $\tau$ and $C_{\mathrm{R}}$ take the forms
 \begin{align}
     C_{\mathrm{R}} &= \frac{9 e^{5/2} B^{3/2} \sin{\psi}}{4\sqrt{\nuobs \me c}}, \\
      \tau &= \sqrt{\frac{3 e B}{16 \nuobs \me c}} \left(\frac{E}{\me c^2}+1\right),
 \end{align}
 where $e$ is the electron charge, $\nuobs$ is the observing frequency, and $\psi$ is the pitch angle with the magnetic field. Different values of $\psi$ are used and discussed in Sections \ref{sec::1D} and \ref{sec::3D}.

 Before moving on, we briefly discuss Eq. \ref{eq::dPdVdv}. If Eq. \ref{eq::dPdVdv} is integrated over a entire downstream width/cooling region, the synchrotron spectral index matches the one in Eq. \ref{eq::ainj_mach} and it is computed from the energy spectral index, Eq. \ref{eq::s_mach}, as $\aint = s/2$. Furthermore, we want to stress that $\xi_{\mathrm{e}}$ (the acceleration efficiency in Eq. \ref{eq::cspec}) is not the only term in Eq. \ref{eq::dPdVdv} that determines the amount of radio emission produced by a shock with a given Mach number. The analytical integration of Eq. \ref{eq::dPdVdv} results in a function that rapidly decreases for low Mach number shocks\footnote{Compare with the function $\Psi(M)$ and figure 4 given in \citet{2007MNRAS.375...77H}.}, i.e. $M \lesssim 3$, making clear its strong dependence on the Mach number. For a detailed discussion, we refer to \citet[][see section 2.5 and figure 4]{2007MNRAS.375...77H} and \citet{2020A&A...634A..64B}.

 \subsection{Mach numbers and their distribution}\label{ssec::distributions}
To analyse the Mach number discrepancy in cosmological simulations, we produced mock observations. Here, we measured the radio and X-ray Mach number as similar to observations as possible. Mach numbers were computed from the radio spectral index, i.e. Eq. \ref{eq::ainj_mach}, and the temperature/surface brightness jump observed in X-rays, see Section \ref{ssec::xray_mach_number}. We refer to these two measurements as the \textit{radio} and \textit{X-ray} Mach number, respectively.

The simulations demonstrate that the corresponding merger shock fronts show a wide distribution of Mach numbers. We introduce three different weighting schemes to characterise the Mach number distributions, that produce the radio relics, and to compute their averages: (1) \textit{volume-weighted}, where all Mach numbers were weighted equally; (2) \textit{radio-weighted}, where we used the total radio power produced by each Mach number, $P_{\radio}(M)$, as a weight, and  (3) \textit{X-ray weighted}, where the X-ray emission at the shock front, $L_{\mathrm{X}(\mathrm{cell)}}$ (see Sec. \ref{ssec::enzo}), was used as a weight. The differently weighted means of each distribution were computed as
 \begin{align}
 \langle M \rangle_{w} = \frac{\sum\limits_{i = 1}^{N_{\mathrm{cells}}} w_i M_i}{\sum\limits_{i = 1}^{N_{\mathrm{cells}}} w_i}, 
 \label{eq::average}
\end{align}
here, the sum was taken over all simulation cells that contribute to the shock front and, hence, to the radio relic emission. In Eq. \ref{eq::average}, $w_i$ is the weight of each distribution: 1) $w_i = 1$, for \textit{volume-weighted}; 2) $w_i = P_{\radio}(M)$ for \textit{radio-weighted}; 3) $w_i = L_{\mathrm{X}(\mathrm{cell)}}$ for \textit{X-ray weighted}. In the following, we refer to these averages as volume-, radio- and X-ray-weighted average Mach number.
 \section{Synthetic shock fronts}\label{sec::1D}
 
 Before investigating the Mach number discrenpancy in 3D cosmological simulations, we take a step back and investigate the Mach number discrenpancy for synthetic shock fronts. We constructed these synthetic shock fronts by fixing all input parameters,  e.g. magnetic field or gas temperature, except the Mach number. This approach provided us with a controlled environment, as we could control the input Mach number distribution.
  
\begin{figure}
  \includegraphics[width = 0.49\textwidth]{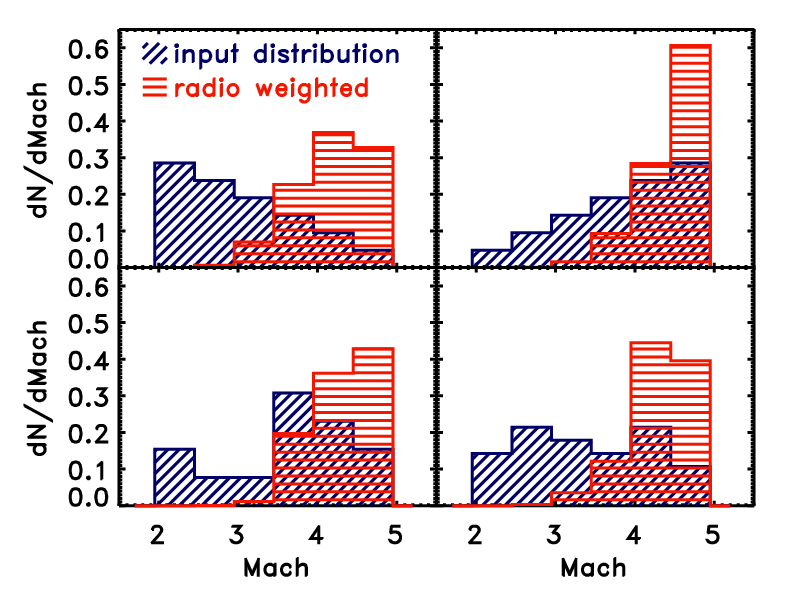} \\
    \includegraphics[width = 0.49\textwidth]{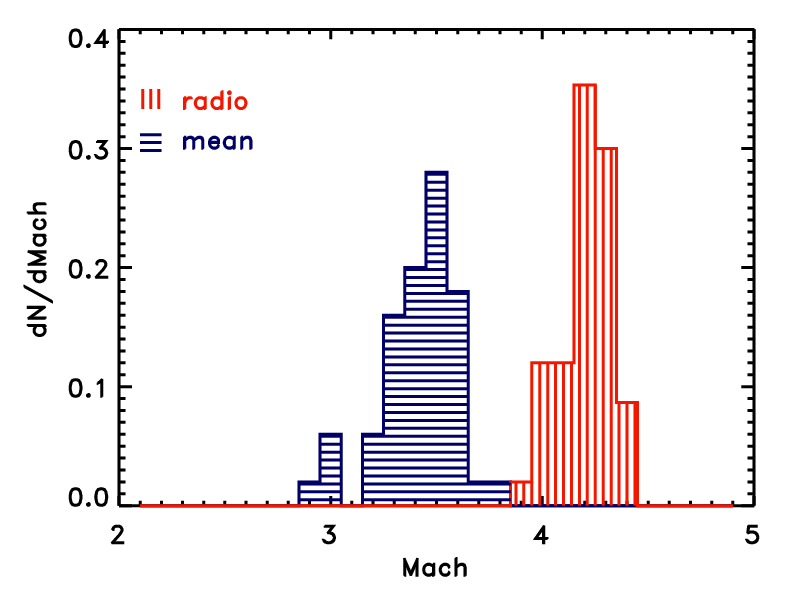} 
  \caption{Synthetic shock fronts: Mach number distributions. Top: Examples of input Mach number distributions, that were used to build the synthetic shock fronts (blue). The red distributions give the corresponding radio weighted Mach number distributions. Bottom: Comparison of the means  of the input distributions (blue) to measured radio Mach number (red).}
  \label{fig::mach_dist_1D}
 \end{figure}

 We constructed 50 synthetic shock fronts, that consist of Mach numbers with values of $M = \left[2.2, \ 2.7, \ 3.2, \ 3.7, \ 4.2, \ 4.7 \right]$. For each shock, we constructed Mach number distribution across the shock front using these values. For two distributions, we used staircase functions that increase and decrease towards the higher Mach numbers. We estimated the other 48 distributions by randomly drawing a value between one and six from a normal distribution. As an example, we show four distributions in Fig. \ref{fig::mach_dist_1D}.
 
 At $\nuobs = [0.14, 1.4, 5.0] \ \GHz$, we computed the total radio power produced by each synthetic shock. We calculated the radio power up to a distance of $500 \ \kpc$, using  a spatial resolution of $\Delta x = 0.01 \ \kpc$ to ensure the numerical convergence of our procedure (see App. \ref{app::conv}).  Here, we assumed an Abell 115-like cluster \citep[see table 1 in][]{2007MNRAS.375...77H}, i.e. using a redshift of $z = 0.1971$ and a downstream temperature of $T = 5 \ \keV$. Furthermore, we used a constant magnetic field strength of $B = 2 \ \mu \mathrm{G}$, a constant acceleration efficiency of $\xi_{\mathrm{e}} = 0.03$ and a constant pitch angle of $\psi = \pi / 2$.
 
 Next, we estimated the integrated radio spectrum and the corresponding radio Mach number for each synthetic shock. In Fig. \ref{fig::mach_dist_1D}, we compare the averages of the input distributions to the estimated radio Mach numbers. For all distributions, the radio Mach number is larger than the corresponding average. Independent of a distribution's shape, the radio Mach numbers are all biased towards the high values of the Mach number distribution. Also for low average Mach numbers, the radio Mach numbers are closer to the highest values of the underlying distribution, i.e. $\ge 3.7$.
 
\begin{figure*}
   \includegraphics[width = 0.3375\textwidth]{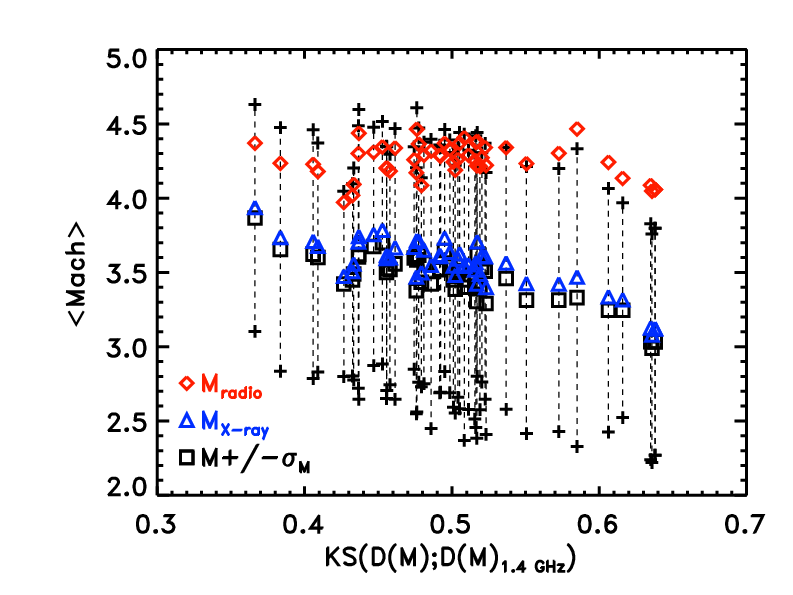}
   \includegraphics[width = 0.325\textwidth]{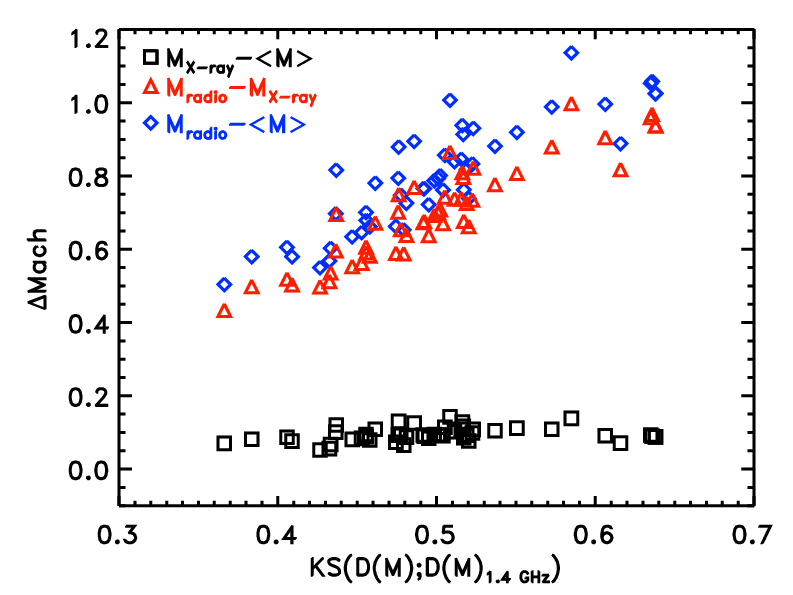}
   \includegraphics[width = 0.325\textwidth]{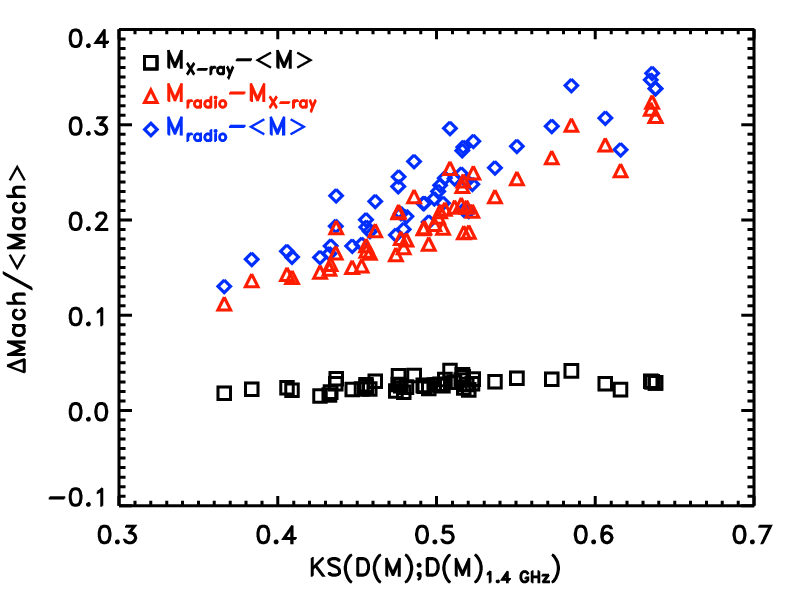}
   \caption{Synthetic shock fronts: Comparison of various Mach number measurements to the results of the KS-tests that compare the input Mach number distributions and the radio weighted Mach number distributions. Left panel: Average Mach number plus/minus one standard deviation (black symbols and line) plotted against the results of the KS-test. The red diamonds show the radio Mach numbers and the blue triangles give the X-ray Mach numbers. Middle and right panel: Absolute (middle) and relative (right) differences between different Mach number proxies against the results of the KS-test. The black squares give the differences between the X-ray Mach numbers and the average Mach numbers. The blue diamonds give the differences between the radio Mach numbers and the average Mach numbers. The red triangles give the differences between the radio and X-ray Mach numbers. In all panels, the KS-test was used to compare the Mach number distribution and the corresponding radio weighted Mach number distribution. Some examples for different distributions are given in Fig. \protect{\ref{fig::mach_dist_1D}.}}
   \label{fig::1d_results}
  \end{figure*}

 For each synthetic shock, we computed the corresponding radio weighted distribution, using Eq. \ref{eq::average}, see Fig. \ref{fig::mach_dist_1D}. As the high Mach numbers produce most of the radio emission, the radio weighted distributions is always skewed to high values. Using a Kolmogorov-Smirnov test \citep[KS-Test,][]{Smirnov:1939:EDB}, we compared the input distributions to the radio weighted distributions. Fig. \ref{fig::1d_results} shows the comparison of different Mach number estimates plotted against the results of the KS-test.
 
 First, we compared the average of the input Mach number distribution to the radio and X-ray Mach number. The latter was derived from the temperature jump, using the Rankine-Hugoniot jump conditions \citep[e.g.][and Section \ref{ssec::xray_mach_number}]{Markevitch:2007}. For larger KS-values, the difference between the average Mach number and the radio Mach number increases. The difference is below one standard deviation of the Mach number distribution for smaller KS-values and it is slightly larger than one standard deviation for larger KS-values. On the other hand, the X-ray Mach number always remains fairly close the average Mach number. Finally, both the absolute and relative differences between the X-ray and radio Mach number increase for larger KS-values.
  
 These results show that the radio Mach number is biased towards the high Mach numbers that produce most of the radio emission. On the other hand, the X-ray Mach number appears to be a good proxy for the mean of the Mach number distribution. These findings suggest that the X-ray and radio Mach number characterise two different parts of the Mach number distribution and, hence, the Mach number discrepancy appears naturally. Simulations of relics confirmed that the Mach number distributions of shocks peak at low values that are smaller than the inferred radio Mach number \citep[][and Section \ref{sec::3D}]{2013ApJ...765...21S,wittor2019pol,Dominguez_2020_relicsI}.
  
 Both observations and simulations show that the integrated radio spectrum is dominated by the few high Mach numbers that produce most of the radio power. Now, one can ask, if a Mach number distribution exists, which produces a radio spectrum that mirrors the low Mach numbers of the underlying distribution? For a constant acceleration efficiency, this can only be achieved if the number of low Mach numbers is significantly larger than the amount of high Mach numbers. Hence, we conducted a simple experiment.
 
 We constructed a synthetic shock front that only consist of Mach numbers with $M = 2.2$ and $M = 4.7$. We fixed the number of $M = 4.7$ shocks to one and we continuously increased the number of $M = 2.2$ shocks. For each synthetic shock front, we computed integrated radio spectrum. We estimated the required number of $M=2.2$ shocks, to overcome the $M=4.7$ shock. In other words, we computed how many $M=2.2$ shocks are needed to observe an integrated spectrum with $\aint \approx 1.52$ instead of $\aint \approx 1.09$, i.e. a radio Mach number of $M(\aint)=2.2$ instead of $M(\aint) = 4.7$. We found that about $\sim 10^5$ $M = 2.2$ shocks are required for the radio Mach number to differ by no more than $10 \ \%$ from $M = 2.2$. If the relative difference is supposed to be below $5 \ \%$, at least $\sim 2.3 \cdot 10^5$ $M = 2.2$ shocks are needed. However, the occurrence of these configurations seems unlikely to happen. The number of required weak shocks can be reduced, if they live inside a region with a stronger magnetic field than the strong shocks. Yet, there is no physical motivation for this scenario. Alternatively, one could require a larger acceleration efficiency for weak shocks. Yet, the required efficiencies would still be unrealistically large. We discuss the dependence of the Mach number discrepancy on variations of the magnetic field and acceleration efficiencies in App. \ref{app::conv}.

\section{3D Modelling}\label{sec::3D}
  
 To investigate the spectral properties of radio relics in a more realistic environment, we analysed three simulated clusters that belong to the \textit{San Pedro}-cluster catalogue. The \textit{San Pedro}-cluster catalogue is a continuously growing database of simulated radio relics, which is meant to complement and support observations. Additionally, it has been used to study the shock acceleration of cosmic-ray protons and the role of the obliquity in the ICM \citep{wittor2020gammas,Banfi2020}.
  
 \subsection{\enzo-simulations}\label{ssec::enzo}
 
 The simulations were performed with the cosmological magneto-hydrodynamical code \enzo \ \citep{ENZO_2014}. The cosmological parameters of the simulations are in agreement with the latest results from the Planck-Collaboration \citep[][]{PlanckVI2018}: $H_0 = 67.66 \ \km \, \sek^{-1} \, \Mpc^{-1}$, $\Omega_{\Lambda} = 0.69$, $\Omega_{\mathrm{m}} = 0.31$, $\Omega_{\mathrm{b}} = 0.05$.
  
 Each simulation covers a root grid size of $(140 \ \Mpc / h)^3$ and is sampled with $256^3$ grid cells. To achieve a high and uniform resolution in the clusters' region, we initialised each simulation with five levels of nested refinements and we added one additional layer of adaptive mesh refinement (AMR) to reach a total of $2^6$ refinements. The nested grids were initialised with \music \ \citep{music} and they cover regions that are at least $3.5^3$ times larger than the volume inside $r_{200}^3$, i.e. sizes of $\sim (4.4 \ \Mpc/h)^3 - (6.6 \ \Mpc/h)^3$. The AMR grids cover $\sim (4.7  \ \Mpc/h)^3$ and they are focused on the center of each cluster. To obtain uniform a refinement over the entire AMR region, we enforced the refinement in each cell using the \textit{MustRefineRegion}-criteria implemented in \enzo. On the most refined level, the resolution is $\Delta x \approx 8.54 \ \kpc/h$. In each simulation, we initialised a uniform magnetic field with $10^{-7} \ \mathrm{G}$ in each direction.
   
 In this work, we analysed three clusters from the \textit{San Pedro}-cluster catalogue (namely, \textit{SP2m}, \textit{SP3r} and \textit{SP8m}) with masses $M_{500}\approx (3-6) \cdot 10^{14} \ \Msun$. They have been selected for this study as they all undergo a major merger which induces large-scale shock waves that produces powerful radio relics. The mergers' mass ratios are: $1:3.9$, $1:2.4$ and $1:1.8$.
 
 Following \citet{Vazza:2009_shock}, we used a velocity jump method to identify shock waves. The velocity jump shock finder uses the 3-dimensional velocity information to compute the Mach numbers along the three coordinate axes. The final shock strength is computed as $M = \sqrt{M_x^2 + M_y^2 + M_z^2}$. In our analysis, we only included shocks with a strength above $M \geq 1.5$, as lower values might be of numerical nature. In each shocked cell, the three components of the Mach number point from the post-shock to the pre-shock and, hence, they give the shock's propagation direction and the shock normal $\mathbf{n}_{\mathrm{shock}} = \left(M_x,M_y,M_z\right)^{\mathsf{T}}$.
 
 Using the model described in Section \ref{sec::equations}, we computed the associated radio relic emission of the detected shock waves. Therefore, we attached a radio profile (see Eq. \ref{eq::dPdVdv}) that extends opposite to the shock normal into the downstream, to each shocked cell. We evaluated Eq. \ref{eq::dPdVdv} at distinct grid points in the downstream region, using a resolution of $\Delta x = 1 \ \kpc$. We calculated the radio profiles to the point where the local radio power drops below $10^{20} \ \erg \, \sek^{-1} \, \Hz^{-1}$. This is way below what is observable. However, it ensures that we did not miss any flux in our computation. We assigned the emission to the grid cell that hosts the grid point, assuming the shock surface matches the cell surfaces. For the further analysis, we only used those grid cells that fulfilled the following two conditions: its emissivity on the three-dimensional grid has to be non-zero and the grid cell has to contribute to a bright pixel in the radio map with a luminosity above $10^{25} \ \erg \, \sek^{-1} \, \Hz^{-1}$, i.e. the sum of the emission of all contributing grid cells must be larger than that value.
 
 For the pitch angle $\psi$, we used the angle between the velocity field and the magnetic field in the shocked cells. Here, we assumed an acceleration efficiency of $\xi_e = 0.02$ \citep[e.g.][]{2007MNRAS.375...77H,wittor2019pol}. The acceleration efficiencies of shocks in the ICM are still loosely constrained and, in principle, they could be a function of the Mach number. In App. \ref{sssec::acceleff}, we tested how the Mach number discrepancy changes if $\xi_e$ is a function of the Mach number. We found that the discrepancies are rather unaffected by the chosen acceleration efficiency. 
 
 We computed the X-ray emissivity with the B-Astrophysical Plasma Emission Code (B-APEC)\footnote{https://heasarc.gsfc.nasa.gov/xanadu/xspec/manual/Models.html}. Assuming a constant temperature and composition in every simulation cell, we computed the X-ray emission, $L_{\mathrm{X}} \propto \rho^2 \Lambda(T,Z) \dd V$, in the energy range $0.5$ to $2 \ \keV$ assuming a metallicity of $0.3$ \citep[with respect to the abundance table given in][]{Anders_Z03}. Furthermore, we calculated the corresponding spectroscopic temperature map following \citet{2004MNRAS.354...10M}:
 \begin{align}
  T_{\mathrm{spec}} = \frac{ \sum (T \cdot T^{-0.75} \cdot \rho^2) }{ \sum (T^{-0.75} \cdot \rho^2)}. \label{eq::tspec}
 \end{align}
 Here, $T$ and $\rho$ are the gas temperature and density. The sum is taken along grid cells that lie along the projection direction. 
  
\begin{figure*}
 \includegraphics[width = 0.33\textwidth]{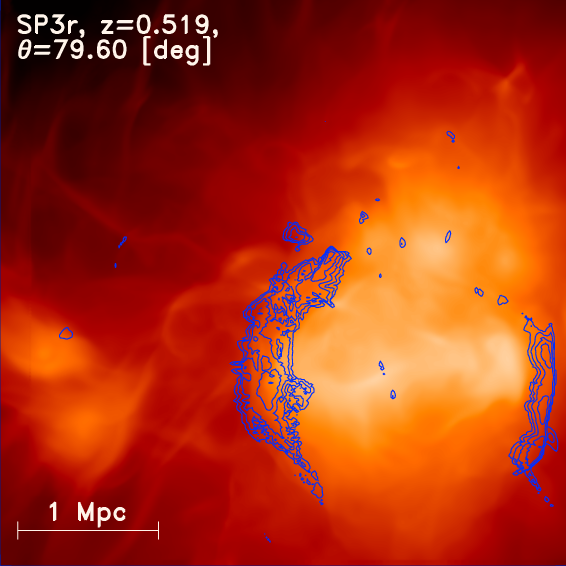} %\\
 \includegraphics[width = 0.33\textwidth]{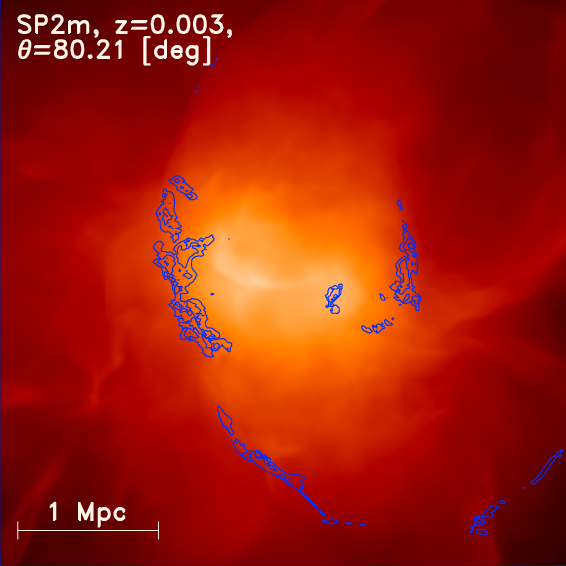}
 \includegraphics[width = 0.33\textwidth]{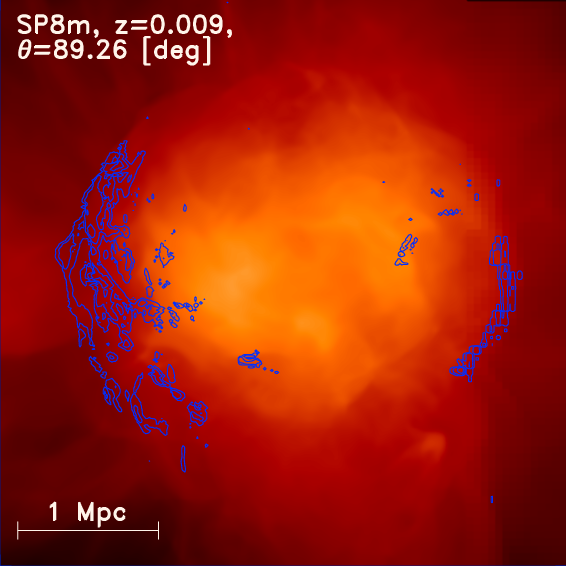}
 \caption{3D analysis: The radio contours at $1.4 \ \GHz$, $[10^{25}, \ 10^{26}, \ \cdots, \ 10^{30} \ \erg \, \sek^{-1} \, \Hz^{-1} \, \mathrm{pixel}^{-1}]$, on top of the X-ray luminosity of three relics from our sample. In each panel, the bar marks a length of $1 \ \Mpc$ comoving. The label in each panel indicates the orientation of the relic with respect to the observer and its redshift.}
 \label{fig::cluster_sample}  
\end{figure*}

\begin{figure}
 \includegraphics[width = 0.49\textwidth]{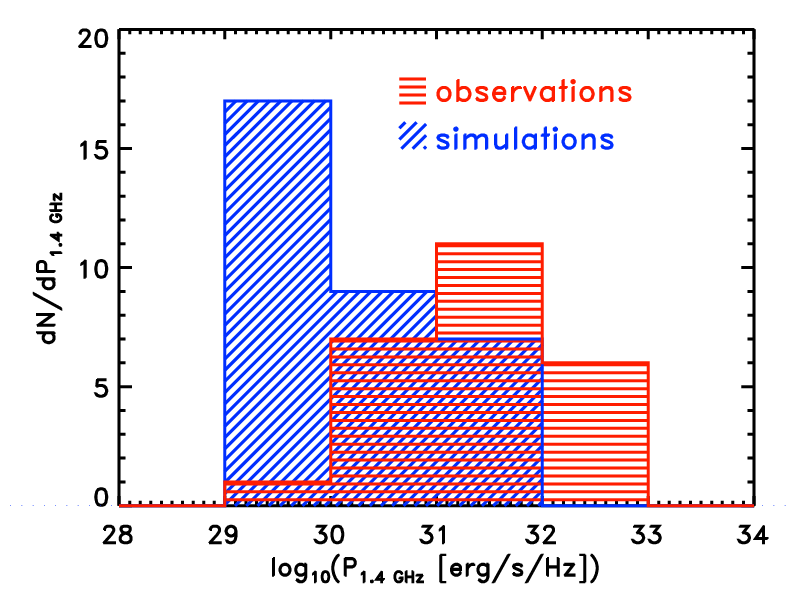} \\
 \includegraphics[width = 0.49\textwidth]{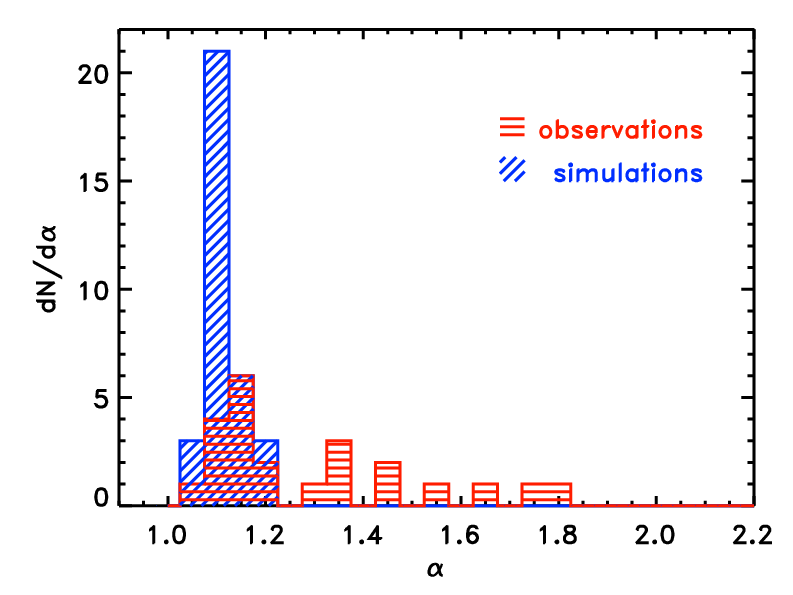} 
 \caption{3D analysis: Comparison of the distributions of the radio power (top) and the integrated spectral indices (bottom). The blue histograms present the data taken from the simulations and the red histograms show the observations, see Table \protect{\ref{tab::observations}}.}
 \label{fig::comparison}
\end{figure}

 \subsection{A sample of simulated radio relics} \label{ssec::results_sample}

 Using the methods described in Sections \ref{sec::equations} and \ref{ssec::enzo}, we modelled the radio relics that were identified by eye in the three simulated galaxy clusters. Each relic is visible for more than one snapshot of the simulation. Hence, we computed the radio emission of the relics found in the clusters SP2m and SP8m at three different times. For the brightest relic of our sample, the one found in cluster SP3r, we calculated the radio emission at five different timesteps. We analysed each relic seen along the three different coordinate axes. Strictly speaking, relics, that are produced by the same shock wave and that are seen in a different projection, are not independent objects. However, we want to emphasise that we treated them as such. In Fig. \ref{fig::cluster_sample}, we show three different relics of our sample. In App. \ref{app::clusters}, we provide the maps of the other relics and take a closer look at the gas properties of each individual cluster. 
  
 Each relic has a smaller counterpart. However, the counter relics in the cluster SP3r and SP8m, i.e. left and right panel of Fig. \ref{fig::cluster_sample}, are too close to the edge of the simulation box for an individual analysis. For the smaller counter relic in cluster SP2m, i.e. middle panel of Fig. \ref{fig::cluster_sample}, we did not detect any X-ray Mach number and, therefore, we did not include it in our sample. In nature, one cannot distinguish, whether a relic is a continuous structure along the line-of-sight or if it consists of two relics projected on top of each other. Hence, the counter relics still contribute to the radio emission of the main relic, when they are a seen in projection.
  
 In the following, we compare the Mach numbers measurements using a variety of proxies. To avoid confusion in the terminology, we want to recall how we name the different proxies. The \textit{radio Mach number} was derived from the integrated radio spectrum measured for each relic (see Eq. \ref{eq::ainj_mach}).  The \textit{density jump based X-ray Mach number} and \textit{temperature jump based X-ray Mach number} were computed from the density jump and temperature jump observed in the X-ray surface brightness distribution (see Section \ref{ssec::xray_mach_number}). Furthermore, we compare the estimates to the underlying Mach number distributions and their average values (see Section \ref{ssec::distributions}). When computing the \textit{Mach number distribution} or \textit{volume weighted Mach number distribution}, the Mach number obtained for each simulation cell was weighted equally. The average of the volume weighted Mach number distribution is called \textit{volume weighted average Mach number}. A \textit{radio weighted Mach number distribution} used the total radio power produced by each Mach number as a weight. \textit{X-ray weighted Mach number distribution} refers to distributions that were computed using the X-ray luminosity of the grid cell that is occupied by the shock/Mach number as a weight. The corresponding averages are named \textit{X-ray weighted average Mach number} and \textit{radio weighted average Mach number}. 
  
  \subsubsection{Radio power and radio Mach number}\label{ssec::radio_mach}
  
  The radio powers at $1.4 \ \GHz$ of the simulated relics all lie in the range $P_{1.4 \ \GHz} \approx (0.1-20) \cdot 10^{30} \ \erg \, \sek^{-1} \, \Hz^{-1}$. In Fig. \ref{fig::comparison}, we compare the radio power and the integrated spectral index of the simulated sample to the ones of the observed sample (see Table \ref{tab::observations}). However, not all observational works report the total relic power. Hence, we converted the given fluxes using 
 \begin{align}
  P_{1.4  \ \GHz} = \frac{4 \pi \cdot D_L^2 \cdot S_{1.4 \ \GHz}}{(1+z)^{1-\alpha}} \label{fig::p14_from_s14} .
 \end{align}
 To compute the luminosity distance $D_L$, we used the same cosmology as in the simulation and not as quoted in each observational paper. The simulated and observed relics have similar radio powers, i.e.  they are all above $P_{1.4 \ \GHz} = 10^{29} \ \erg \, \sek^{-1} \, \Hz^{-1}$. Though, the simulated sample consists of more relics with low radio power, while the observational sample contains more relics with high radio power. This discrepancy in radio power is often found, as most simulations produce radio powers that are at the lower end of what is observed \citep[e.g.][]{Stuardi2019,wittor2019pol}. This might be based on the lack of fossil electrons in the simulations. A comparison with a larger sample of observed radio relics \citep[e.g.][]{2014MNRAS.444.3130D,2015MNRAS.453.3483D,2017MNRAS.470..240N} suggests that the fainter relics of our sample might not be detectable. On the other hand, studies by \citet{2012MNRAS.420.2006N,2017MNRAS.470..240N} and \citet{Bruggen_2020} predict that a large number of relics exists that lie below the detection limits of current radio telescopes. New observations revealed the existence of such a low-luminous class of radio relics that were previously either not observed or not classified as radio relics \citep{locatelli2020dsa,2020MNRAS.499..404P}.
 
\begin{figure}
 \includegraphics[width = 0.49\textwidth]{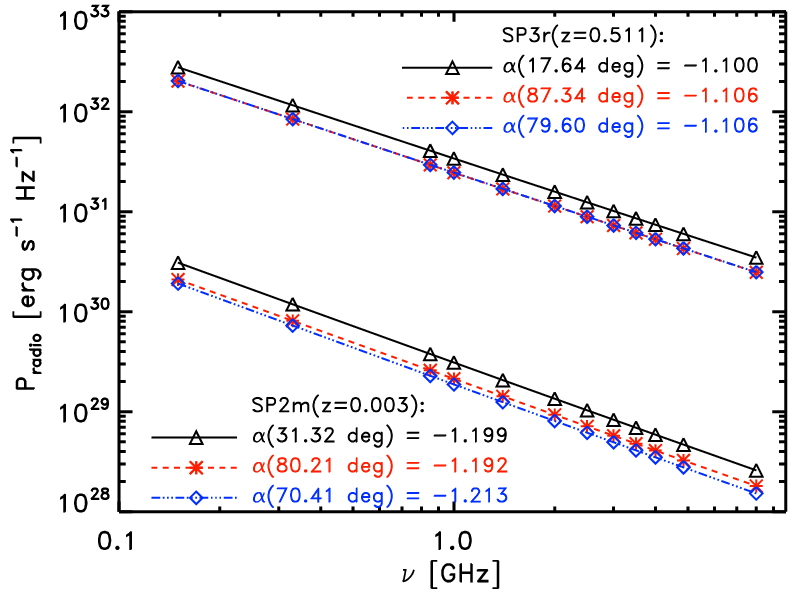}
  \caption{3D analysis: Two examples of integrated radio spectra for the relics found in SP3r at $z = 0.511$ (top lines) and SP2m at $z = 0.003$ (bottom lines). For each relic, we display the integrated radio spectra measured for the three different orientations, see labels.}
  \label{fig::spectra}
\end{figure}

 For each simulated radio relic, we computed the integrated radio spectrum in the range of $150 \ \MHz$ to $4.85 \ \GHz$. As an example, we show the spectra of two relics in Fig. \ref{fig::spectra}. The spectral index distribution of the simulated sample is narrower than the one of the observed sample (see Fig. \ref{fig::comparison}). The integrated spectral indices of the simulated sample are mostly flat and do not exceed $\sim 1.24$, while the observed ones are also steeper than $1.24$. We discuss possibilities for this difference in Section \ref{sec::summary}. In the simulation, the corresponding radio Mach numbers are rather high, i.e. $M_{\radio} \approx 3.13-4.72$. This range of Mach numbers matches very well the top end of what is measured by observations (compare with Table \ref{tab::observations}).

\begin{figure*}
  \includegraphics[width = 0.49\textwidth]{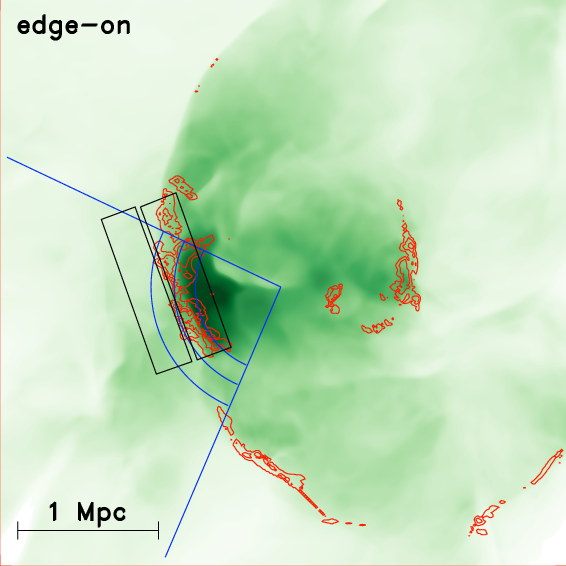}  
  \includegraphics[width = 0.49\textwidth]{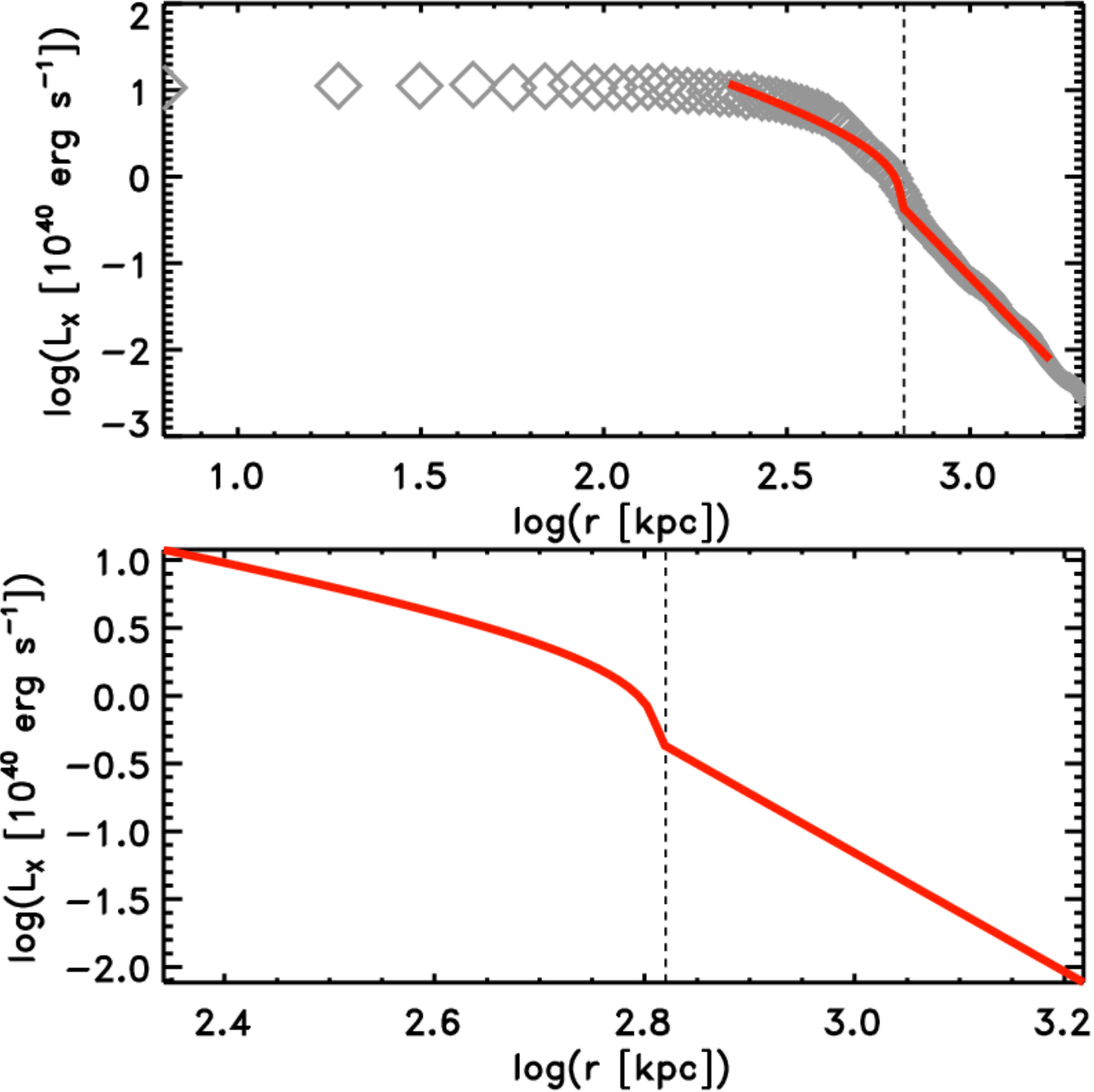}
  \caption{3D analysis: Extraction of the X-ray surface brightness profile. The left panel shows the map of the spectroscopic temperature overplotted with the radio contours at $1.4 \ \GHz$, $[10^{25}, \ 10^{26}, \ \cdots, \ 10^{30} \ \erg \, \sek^{-1} \, \Hz^{-1} \, \mathrm{pixel}^{-1}]$. The black rectangles mark the regions in which the temperature jump was estimated. The blue solid lines show the shape of the circular annuli used to derive the X-ray surface brightness profile. On the right, the top panel shows the surface brightness profile extracted from the cluster centre beyond the relic. The red solid line, marks the region of the fit, which was used the estimate the density jump based X-ray Mach number. The bottom panel shows the fitted profile across the relic. In both panels, the vertical lines is the estimated position of the relic's outer edge.}
  \label{fig::tspec_map}
 \end{figure*} 
 
\subsubsection{X-ray luminosity and X-ray Mach number}\label{ssec::xray_mach_number}
 
 The clusters in our sample have X-ray luminosities in the range of $L_X \approx (1.6-10.1) \cdot 10^{44} \ \erg \, \sek^{-1}$. We followed the method outlined in \citet{Sarazin2016} to compute the X-ray surface brightness profiles and corresponding Mach numbers. As an example, Fig. \ref{fig::tspec_map} illustrates the method for one of our simulated clusters. For each relic, we computed the X-ray surface brightness profile inside circular pie annuli (see blue regions in Fig. \ref{fig::tspec_map}). The annuli are focused at the centre of the projected gas mass and they span angles that depend on the extend of the relic. To each profile, we fitted the functional form given by equation 1-5 in \citet{Sarazin2016}. 
 
 In our analysis, we assumed that the X-ray emissivity is constant within a given annulus, and that, if present, it has a circular discontinuity. On the two sides of the discontinuity, a power-law function was fitted to the surface brightness (see red lines in Fig. \ref{fig::tspec_map}). The radius of the discontinuity is defined from the fit of the broken power-law, and it is where we measured the jump in X-ray emissivity. The observed X-ray emissivity 
 \begin{align}
  \epsilon_{\mathrm{X-ray}} = \rho^2 \Lambda_{\mathrm{obs}}(T,Z)
 \end{align} 
 depends on the local gas density $\rho$ and the observed temperature dependent emission function\footnote{We note that the emission function that is derived from the projected maps is not to be mistaken with the $\Lambda(T,Z)$ used in Section \ref{ssec::enzo} to compute the X-ray emission in the individual grid cells of the simulation.} $\Lambda_{\mathrm{obs}}(T,Z)$. Hence, if $\Lambda_{\mathrm{obs}}(T,Z)$ is known, one can directly compute the compression ratio and corresponding Mach number from the emissivity jump as
 \begin{align}
  \frac{\rho_2}{\rho_1} = \sqrt{\frac{\epsilon_{\mathrm{X-ray},2}}{\epsilon_{\mathrm{X-ray},1}} \frac{\Lambda(T_1,Z)}{\Lambda(T_2,Z)}}.
 \end{align}
 In the regions of interest, we estimated temperature dependent emission functions in the $0.5-2 \ \keV$ band from the spectroscopic temperature map, (see Eq. \ref{eq::tspec}). Here, we assumed a constant metallicity of $Z=0.3$ \citep[with respect to the abundance table given in][]{Anders_Z03} and evaluated the count-rate for a generic response. 
 
 We also determined an X-ray Mach number based on the temperature jump. Therefore, we computed the average spectroscopic temperatures in rectangular regions that lie behind and in front of the above defined position of the discontinuity (see, e.g., black rectangles in Fig. \ref{fig::tspec_map}). The size of the regions have been defined such that the signal of the temperature jump is maximized. Using the measured temperature jump, we computed the X-ray Mach number using the Rankine-Hugoniot jump conditions \cite[e.g.][]{Markevitch:2007}. For relics that do not have the typical arc-shape of edge-on relics, i.e. relics that are rather seen face-on or at an oblique angle, we tried to determine the X-ray Mach numbers in the region in which the shock front is the most prominent.
 
 In the simulation, we measured temperature jump based X-ray Mach numbers for most relics. All of the non-detections occur for relics in SP8m that are seen face-on. Surprisingly, we also detected temperature jump based X-ray Mach numbers for all remaining face-on relics. For these relics, the measured Mach numbers are always very low, i.e. $1.53$ or less. In general, the temperature jump based X-ray Mach numbers of the simulated sample all lie in the range of $M_{\mathrm{X-ray}}(\Delta T) \approx 1.09 - 4.15$. Consequently, they span a broader range than the observed sample, that is $1.27-3.7$ (see Table \ref{tab::observations}).
 
 We only detected a density jump based X-ray Mach number for a fraction of the simulated relics. We found the presence of a density jump for all relics seen face-on in our simulation. Yet, in these cases the measured shocks are very weak, i.e. $\leq 1.1$. Only in the cluster SP2m, we could detect a density jump based X-ray Mach number for the relics that are rather seen edge-on. In the other two clusters, we did not detect any density based Mach numbers. In these cases, either the surface brightness fit did not converge or the estimated density jump was larger than 4, in contrast with the Rankine-Hugoniot jump conditions for an adiabatic index of $\gamma = 5/3$. However, the measured density jump based X-ray Mach numbers for the relics seen edge-on take values between $M_{\mathrm{X-ray}} \approx 1.77-4.86$. Except the outlier of $M_{\mathrm{X-ray}} \approx 4.86$, they are of the same strengths as in the observed sample, i.e. $M_{\mathrm{X-ray}} \approx 1.1-2.42$ (see Table \ref{tab::observations}).
 
 As expected, we found that the measured X-ray Mach number is sensitive to the orientation of the shock front to the observer. This is contrary to the radio Mach number, that is rather independent of the viewing angle. If a relic, that is seen edge-on, is rotated to an oblique angle, projection effects smear out the shock and the observed shock strength appears weaker.
  
\subsubsection{Mach number discrepancy}\label{sssec::discrenpancy}
 
 In Fig. \ref{fig::discrepancy}, we plot the Mach number discrepancies estimated in both simulations and in the observations. The discrepancies found in the simulated clusters resemble the ones found in observations. However, there are some significant differences. In the simulations, all radio Mach numbers are larger than the corresponding temperature jump based X-ray Mach number. This is not the case in the observed sample.
 
 In the observed sample, the radio Mach numbers are either larger or equal to the X-ray Mach numbers that are based on the density jump. In the simulations, we found larger density jump based X-ray Mach numbers for two relics found in SP2m. Here, the differences are $\Delta M \approx -0.19$ and $\Delta M \approx -1.15$. We checked if these two relics have some peculiar properties. However, their other properties do not stand out compared to the other relics.
 
 As in the observations, we found a Mach number discrepancy in the simulations, illustrating a systematic difference between the radio and X-ray Mach number. However, the discrepancies, $\Delta M = M_{\radio} - M_{\mathrm{X-ray}}$, found in the simulation are on average larger, i.e. $\langle \Delta M(T)  \rangle_{\mathrm{sim}} \approx 1.7$ and $\langle \Delta M(\rho)  \rangle_{\mathrm{sim}} \approx 2.4$, than in the observations, i.e. $\langle \Delta M(T)  \rangle_{\mathrm{obs}} \approx 0.8$ and $\langle \Delta M(\rho)  \rangle_{\mathrm{obs}} \approx 1.4$. In the following sections, we investigate if the discrepancy is inherent to various properties of the relics themselves or of the host clusters.
  
\subsubsection{Dependence on the radio power}\label{ssec::radio_power}
 
\begin{figure*}
 \includegraphics[width = 0.33\textwidth]{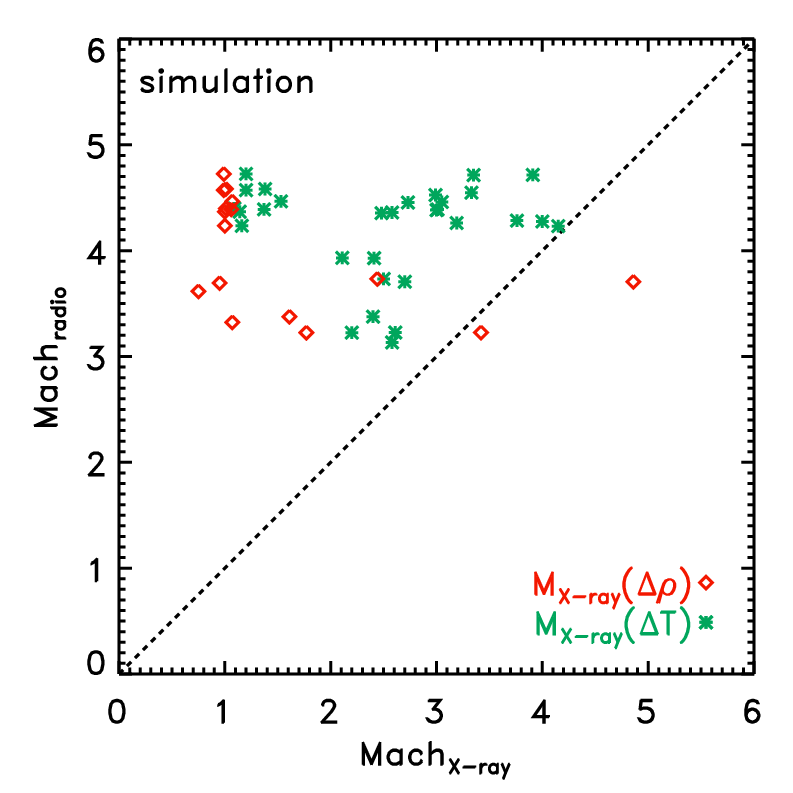}
 \includegraphics[width = 0.33\textwidth]{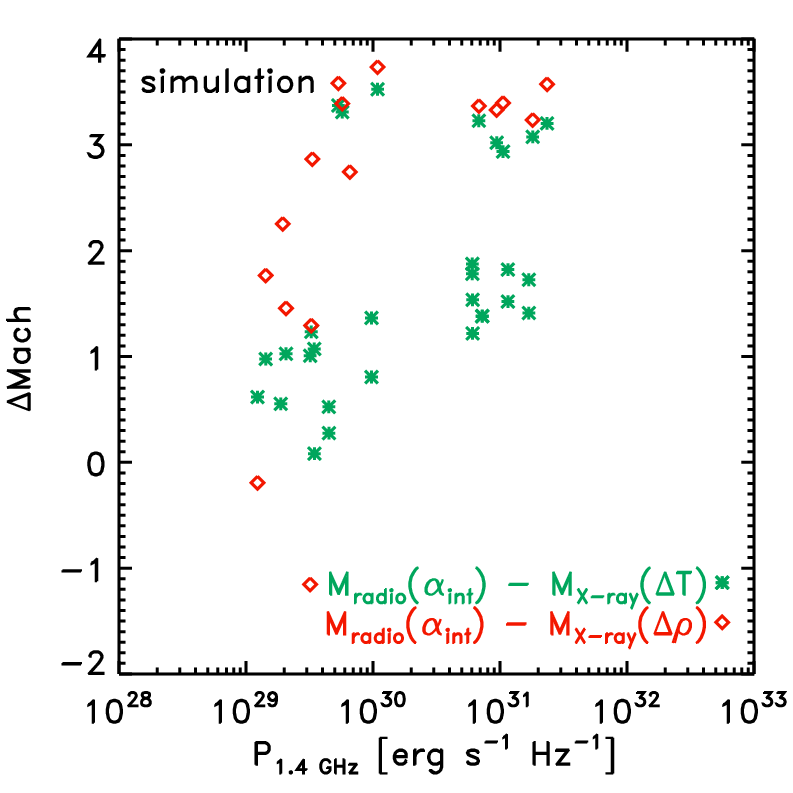}
 \includegraphics[width = 0.33\textwidth]{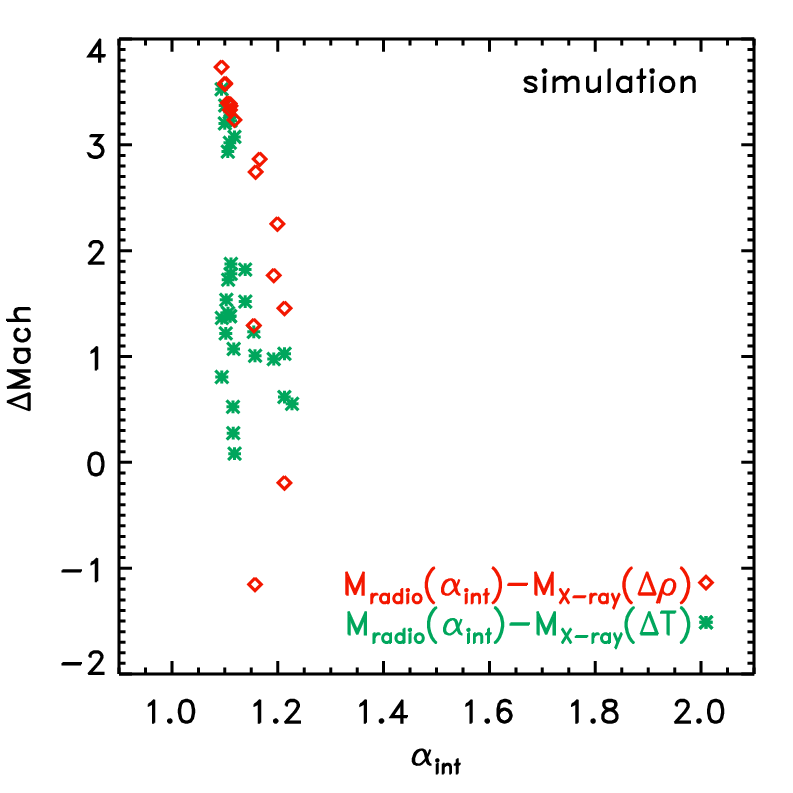} \\
 \includegraphics[width = 0.33\textwidth]{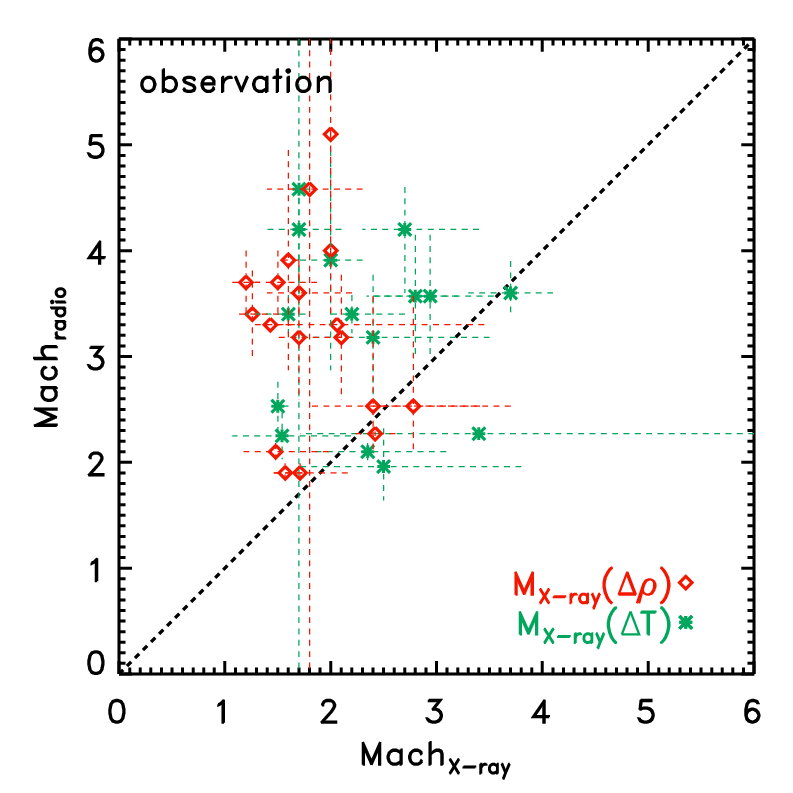}
 \includegraphics[width = 0.33\textwidth]{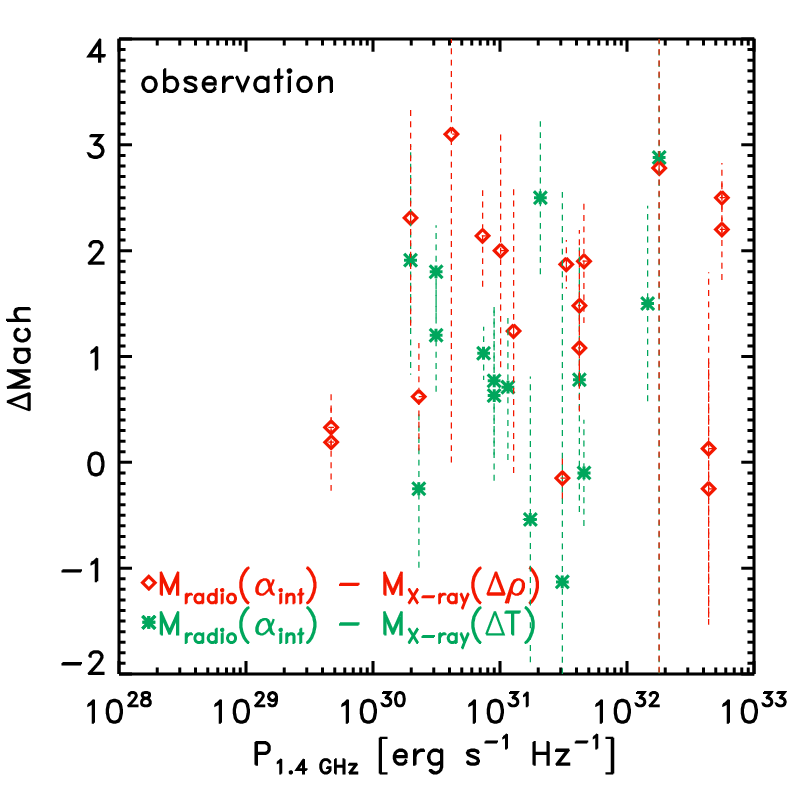}
 \includegraphics[width = 0.33\textwidth]{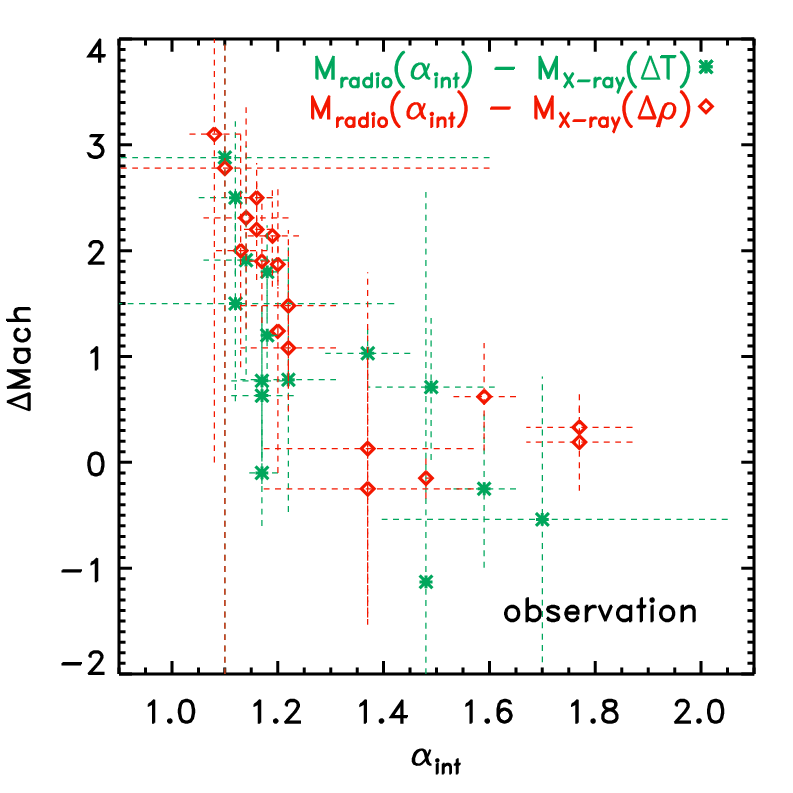} 
 \caption{3D analysis: Mach number discrepancies found in the simulation (top row) and in observations (bottom). The left column displays the radio Mach number plotted against the X-ray Mach number. The middle and right column show the Mach number discrepancy plotted against the radio power at $1.4 \ \GHz$ and the integrated spectral index, respectively. In all plots, the green asterisks refer to the X-ray Mach number measured from the temperature jump. The red diamonds show the results, if the X-ray Mach number is computed from the density jump.}
 \label{fig::discrepancy}
\end{figure*}

 We show in Fig. \ref{fig::discrepancy} the Mach number discrepancy, $\Delta M = M_{\radio} - M_{\mathrm{X-ray}}$, as function of the radio power at $1.4 \ \GHz$ of the relic. In the simulation, we found that the Mach number discrepancy increases for more powerful radio relics. These findings are independent of the method used the measure the X-ray Mach number. In the sample of observed relics, we did not find such a behaviour between the Mach number discrepancy and the radio power. However, this could be due to a selection effect because for a significant number of radio bright relics, such as PSZ1 G108.18-11.53  \citep{2015MNRAS.453.3483D}  or  MACS J1752.0+4440 \citep{2012MNRAS.425L..36V,2012MNRAS.426...40B}, no shocks have been detected in X-rays yet.
 
 Using the same approach, we compared the Mach number discrepancy to the integrated spectral index. Both in the simulations and the observations, there is a strong correlation between the discrepancy and the spectral index, i.e. the discrepancy decreases for steeper spectral indices. This trend is independent of the type of X-ray measurement. However, it is not a surprise, as flatter spectral indices yield larger radio Mach numbers, i.e. Eq. \ref{eq::ainj_mach}, which are not observed in X-ray. 
 
\begin{figure*}
 \includegraphics[width = 0.49\textwidth]{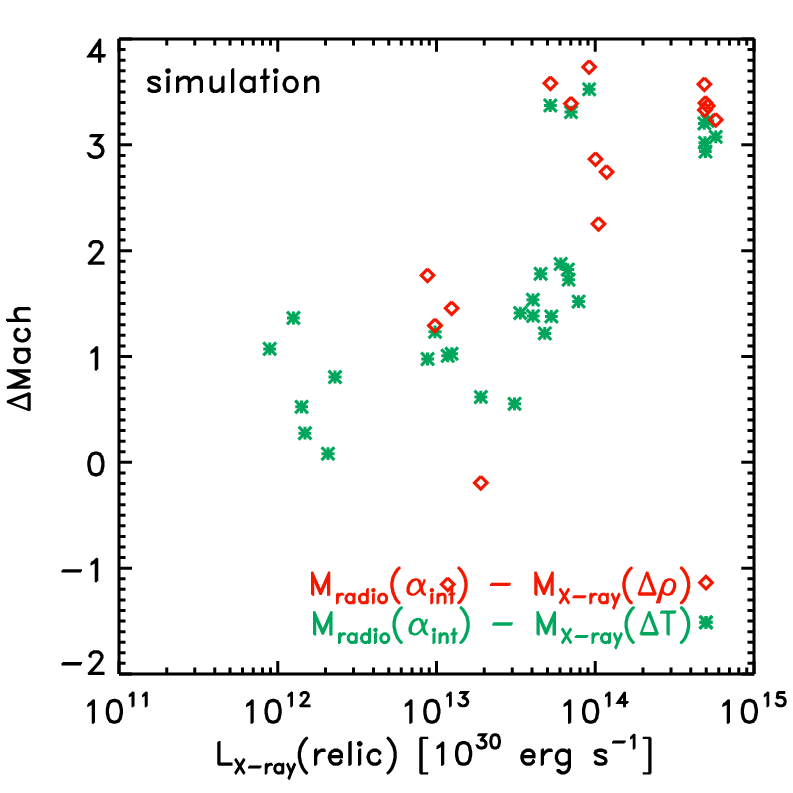} 
 \includegraphics[width = 0.49\textwidth]{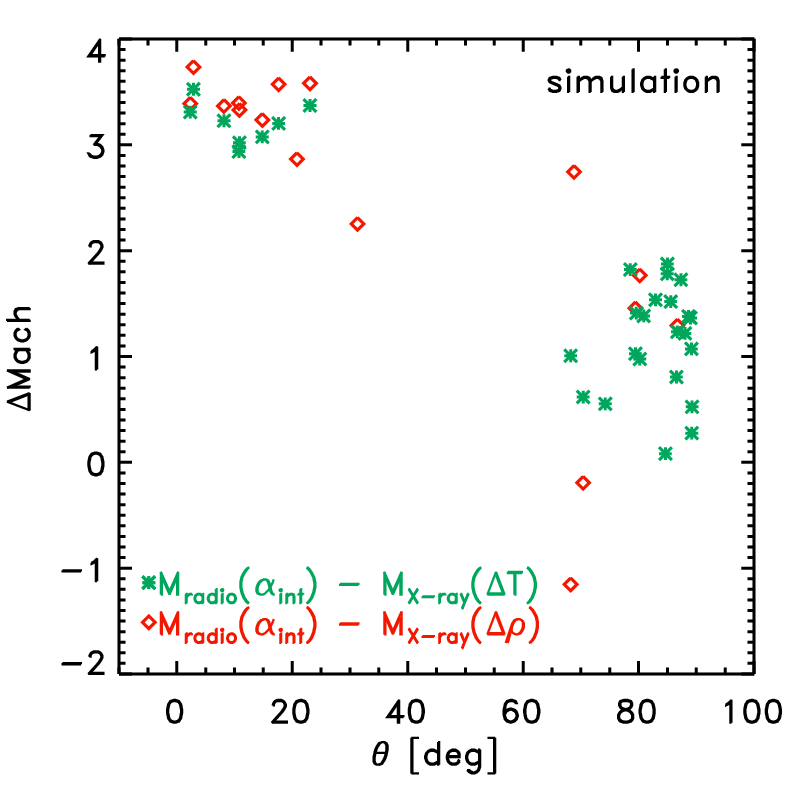}  \\
 \includegraphics[width = 0.49\textwidth]{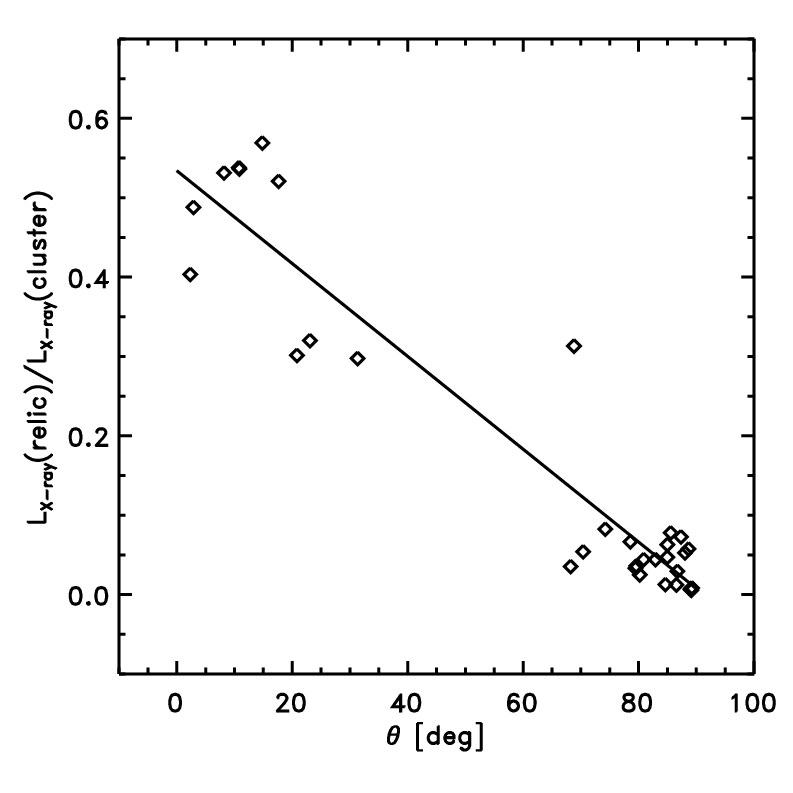} 
 \includegraphics[width = 0.49\textwidth]{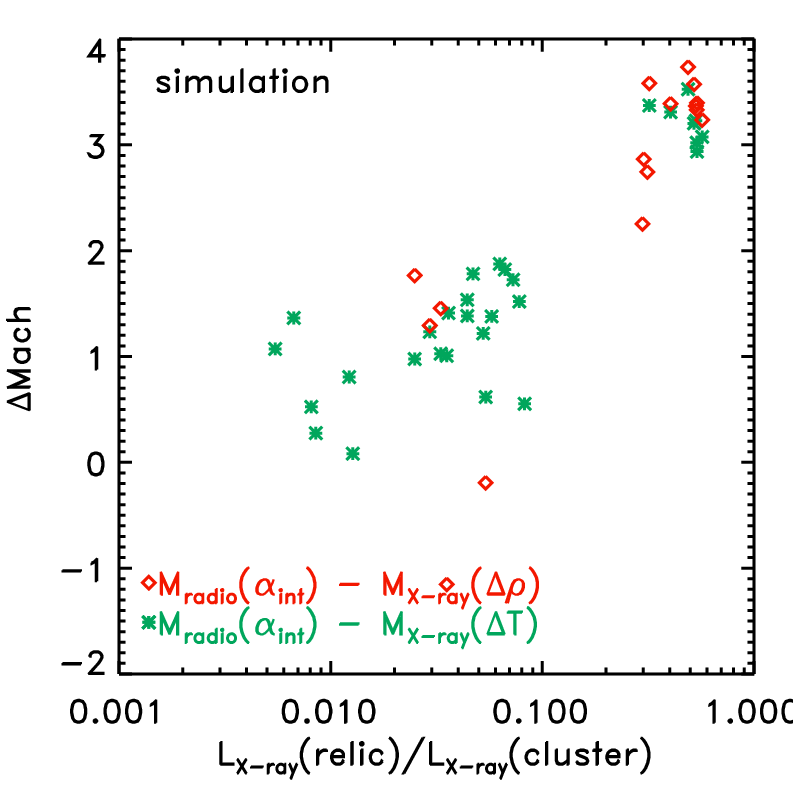} 
 \caption{3D analysis: Mach number discrepancy plotted against properties of the simulated relics. The top left panel displays the Mach number discrepancies plotted against the X-ray luminosity emitted from the regions around the radio relics. The top right panel shows the discrepancy against the viewing angle. Edge-on and face-on relics are seen at angles of $90^{\circ}$ and $0^{\circ}$, respectively. The bottom left panel compares the viewing angle to the X-ray luminosity ratio. The bottom right panel displays the discrepancy plotted against the X-ray luminosity ratio. In all panels, the green asterisk refer to the X-ray Mach number measured from the temperature jump. The red diamonds show the results, if the X-ray Mach number is computed from the density jump.}
 \label{fig::xray_vs_diff_2}
\end{figure*} 
 \subsubsection{Dependence on the X-ray luminosity and viewing angle}\label{sssec::xray_ratio}
 For the simulated relics, we correlated the Mach number discrepancy to the X-ray luminosity of both the host cluster and the volume that is occupied by the relic. No correlation between the X-ray luminosity of the host cluster and the discrepancy was found. Comparing the Mach number discrepancy to the X-ray luminosity in the relics' regions, it can be seen that the discrepancy increases for higher luminosities (see Fig. \ref{fig::xray_vs_diff_2}). As we will discuss below, this trend is connected to the relic's orientation.
 
 Finally, we estimated how the Mach number discrepancy depends on the orientation of the relic to the observer. Therefore, we define the viewing angle, $\theta$, as the angle between the average shock normal and the line-of-sight. If the two are perpendicular to each other, i.e. $\theta \rightarrow 90^{\circ}$, the relic is seen edge-on. Vice versa, if they run parallel, i.e. $\theta \rightarrow 0^{\circ}$, the relic is observed face-on. 
 
 In Fig. \ref{fig::xray_vs_diff_2}, we show the correlation between the Mach discrepancy and the derived angle. If the X-ray Mach number is based on the temperature jump, the discrepancy in Mach number becomes smaller for larger angles. This trend is also observed, if the X-ray Mach number is based on the density jump. However, even if a relic is seen perfectly edge-on, an average discrepancy of $\Delta M \approx 1$ remains.
  
 Measuring the viewing angle is difficult in observations, because only two dimensional information is available. Therefore, we thought of new ways to determine the orientation of the relic. One possibility is to estimate the ratio of the cluster's total projected X-ray luminosity and the X-ray projected luminosity emitted in the relic region. In the following, we will refer to this ratio as the \textit{X-ray luminosity ratio}. 
  
 If the relic is seen face-on, it is most likely projected on top of the cluster's centre, where the X-ray luminosity is the highest. On the other hand, if the relic is seen edge-on, it will most likely be located at the cluster's periphery, where the X-ray luminosity is lower. Consequently, if the X-ray luminosity ratio approaches unity, the relic is most likely seen face-on. The smaller the ratio, the more likely it is that the relic is seen edge-on. Fig. \ref{fig::xray_vs_diff_2} shows that the X-ray luminosity ratio is a trustworthy proxy for a relic's orientation. Relics that are seen edge-on, i.e. $\theta \sim 90^{\circ}$, show a smaller X-ray luminosity ratio than the relics that are observed face-on, i.e. $\theta \sim 0^{\circ}$. Yet, there is still some scatter in the data, as X-ray bright clumps at the relic's location can bias the X-ray luminosity ratio to larger values. We fitted a linear function to the data, i.e. $L_{\mathrm{X-ray}}(\mathrm{relic}) / L_{\mathrm{X-ray}}(\mathrm{cluster}) \approx m \cdot \theta + b$. We found a $b \approx 0.534$ and a slope of $m \approx - 0.0058$, if $\theta$ is measured in degrees, and $m \approx -0.335$, if $\theta$ is measured in radians. Consequently, the X-ray luminosity ratio gives an extra approximated tool to guess a relic's orientation in real observations. 
 
 In Fig. \ref{fig::xray_vs_diff_2}, we plot the Mach number discrepancy against the X-ray luminosity ratio. Similar to the viewing angle, we found that the Mach number discrepancy becomes smaller for smaller ratios. These results also explain why the discrepancy increases for higher X-ray luminosities emitted in the relics' regions: in our set-up, a more luminous relic region indicates a face-on relic and, hence, a relic with a larger Mach number discrepancy.
   
\subsubsection{Connection to the underlying Mach number distribution}\label{sssec::distributions}
 
\begin{figure*}
  \includegraphics[width = 0.49\textwidth]{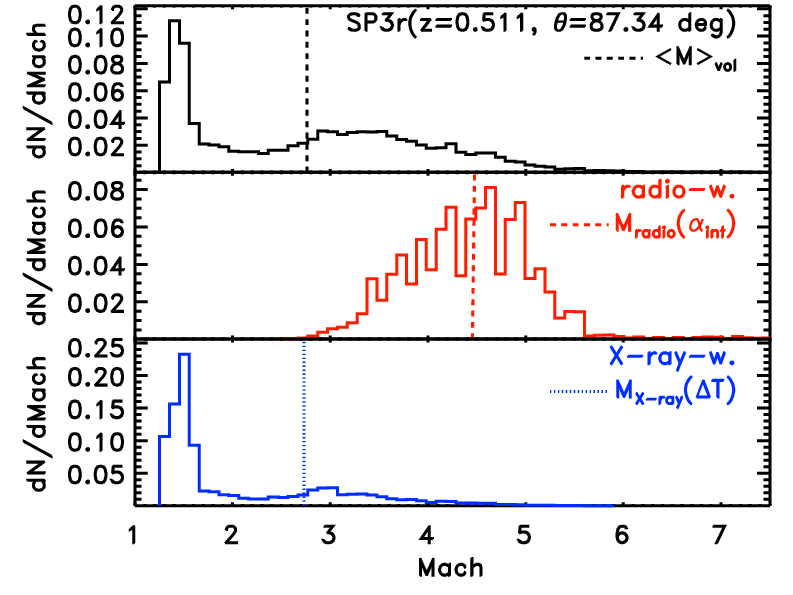} 
  \includegraphics[width = 0.49\textwidth]{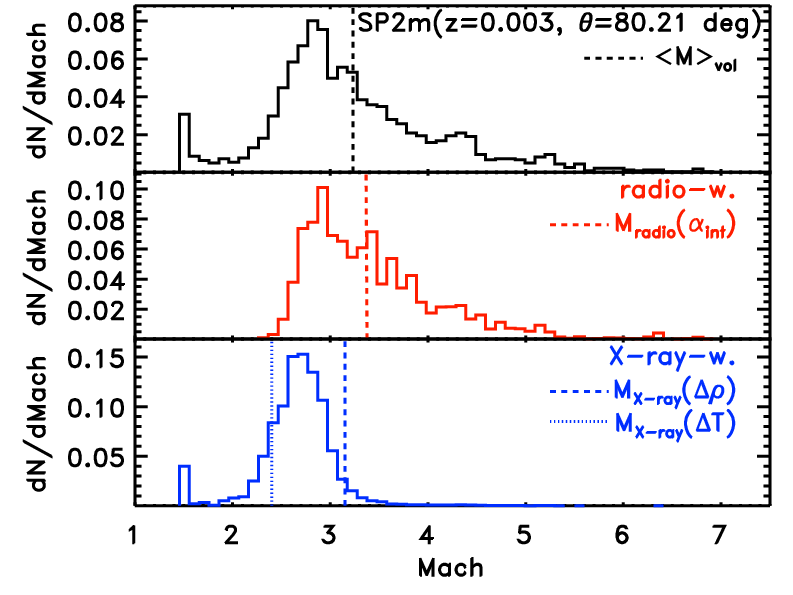}
  \caption{3D analysis: Two examples of 3d Mach number distributions for the relics found in SP3r (left) and SP2m (right panel). In each figure, the three panels give the volume (black, top), radio (red, middle) and X-ray (blue, bottom) weighted Mach number distribution of the shock that produces the relic, measured in 3D. Additionally, we over-plotted with vertical lines the averages of the volume weighted distributions (top) as well as the measured radio (middle) and X-ray (bottom) Mach numbers.}
  \label{fig::sp3r_example}
 \end{figure*}

 Relics do not simply trace a single Mach number but, as simulations indicate, a distribution of Mach numbers. Hence for each relic, we computed the underlying volume weighted, X-ray weighted and radio weighted Mach number distribution. As an example, we plot the three different distributions for two relics in Fig. \ref{fig::sp3r_example}. 
 
 We find that all volume weighted Mach number distributions peak at smaller values. The volume weighted averages all lie in the range of $\langle M \rangle_{\mathrm{vol}} \approx 2.6 - 3.8$ and the corresponding standard deviations are in the range of $\sigma_{M} \approx 0.6 - 1.3$. The volume weighted averages of relics that belong to the same cluster are rather similar. However, this is obvious, because they are the same object either seen from a different projection or at a different timestep. The X-ray weighted averages and radio weighted averages are found in the ranges of $\langle M \rangle_{\mathrm{X-ray}} \approx 2.1 - 3.6$ and $\langle M \rangle_{\mathrm{radio}} \approx 3.1 - 4.7$.
  
 \begin{figure}
 \includegraphics[width = 0.49\textwidth]{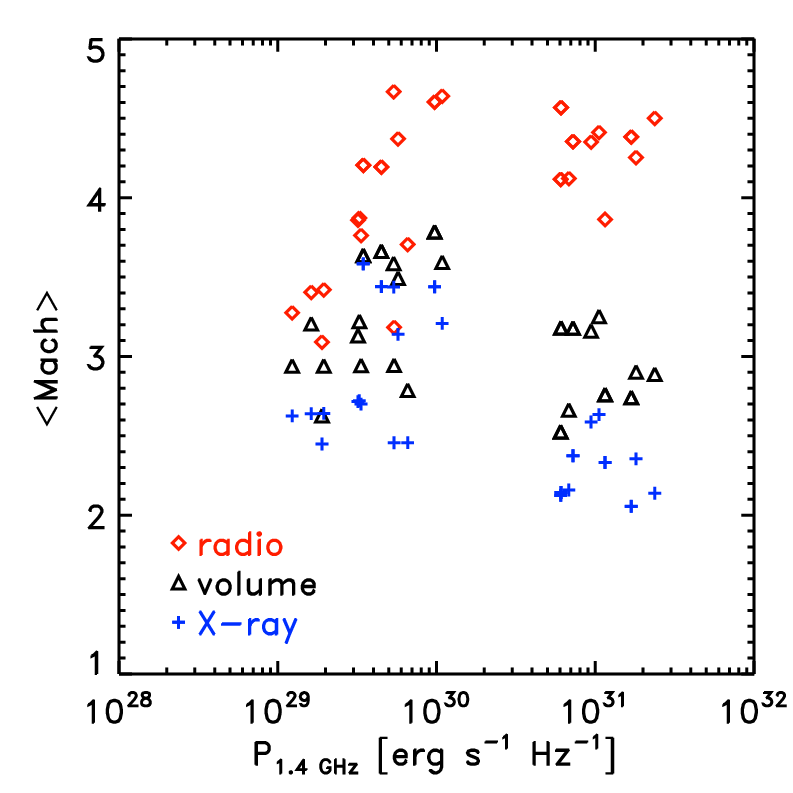}  
 \caption{3D analysis: Averages of the differently weighted Mach number distribution versus the radio power of the relic: volume weighted (black triangles), radio weighted (red diamonds), X-ray weighted (blue plus).}
 \label{fig::main_results}
\end{figure}
  
 In Fig. \ref{fig::main_results}, we plot the average Mach number of each distribution against the radio power at $1.4 \ \GHz$. The differently weighted averages diverge with increasing radio power. The X-ray weighted Mach number slightly decreases for larger radio powers, while the radio weighted Mach number increases. The volume weighted averages remain constant. For all relics, the radio weighted average Mach number is larger than the volume weighted average Mach number. The X-ray weighted Mach number is always smaller than the volume weighted average Mach number. In accordance with the divergence of the averages, a KS-test showed that the differently weighted distributions diverge for larger radio powers.

\begin{figure*}
 \includegraphics[width = 0.49\textwidth]{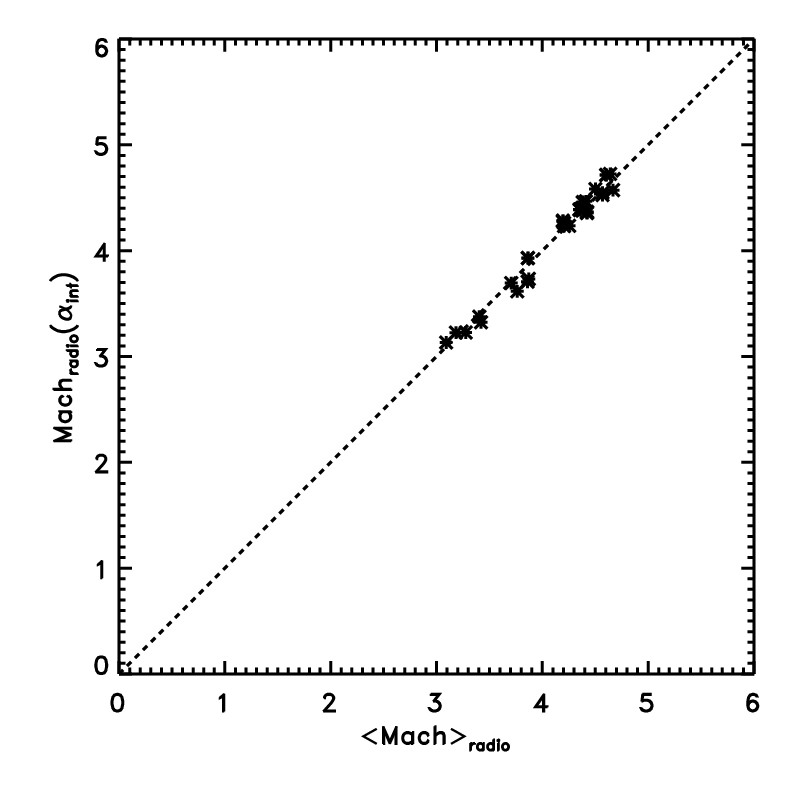} 
 \includegraphics[width = 0.49\textwidth]{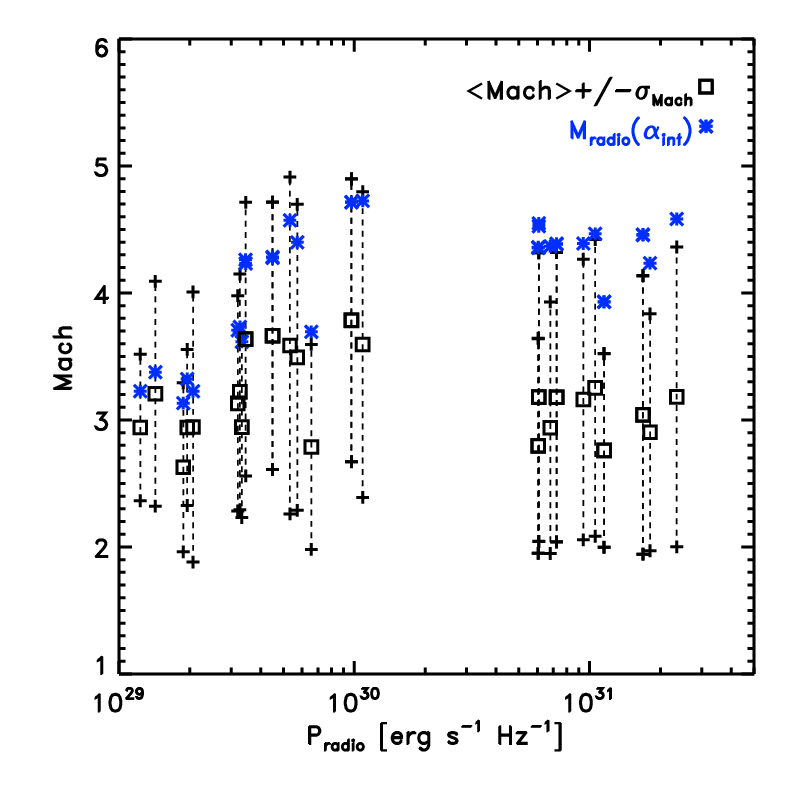}
 \caption{3D analysis, left: Radio weighted Mach numbers plotted against the radio Mach numbers, that were derived from the integrated spectral index. Right: Volume weighted average Mach numbers plus/minus one standard deviation (black lines and symbols) and the radio Mach numbers plotted against the radio power (blue astriks). }
 \label{fig::obs_vs_dist}
\end{figure*} 
  
\begin{figure*}
 \includegraphics[width = 0.49\textwidth]{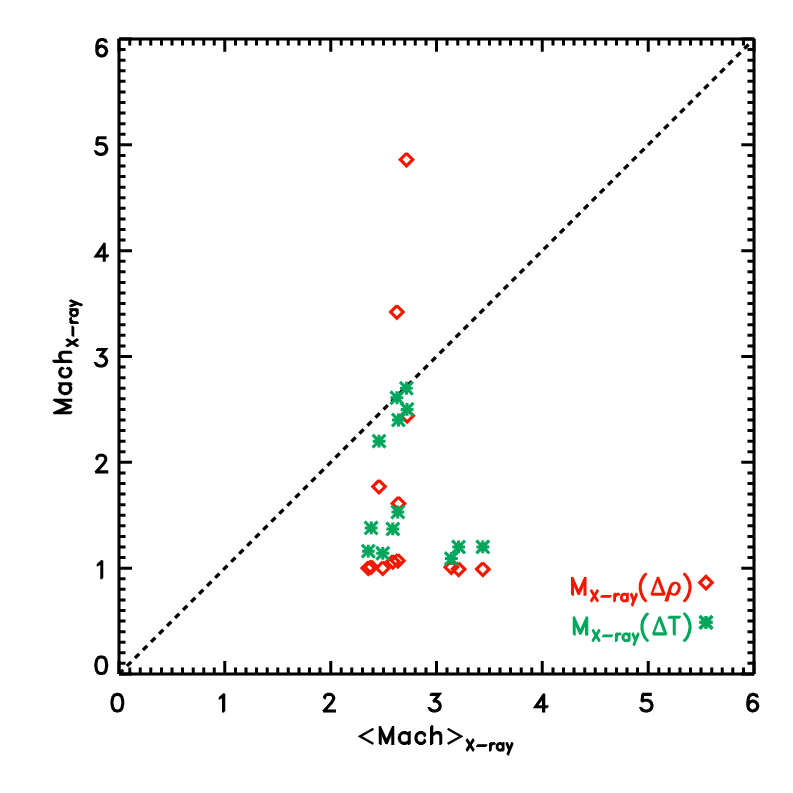}
 \includegraphics[width = 0.49\textwidth]{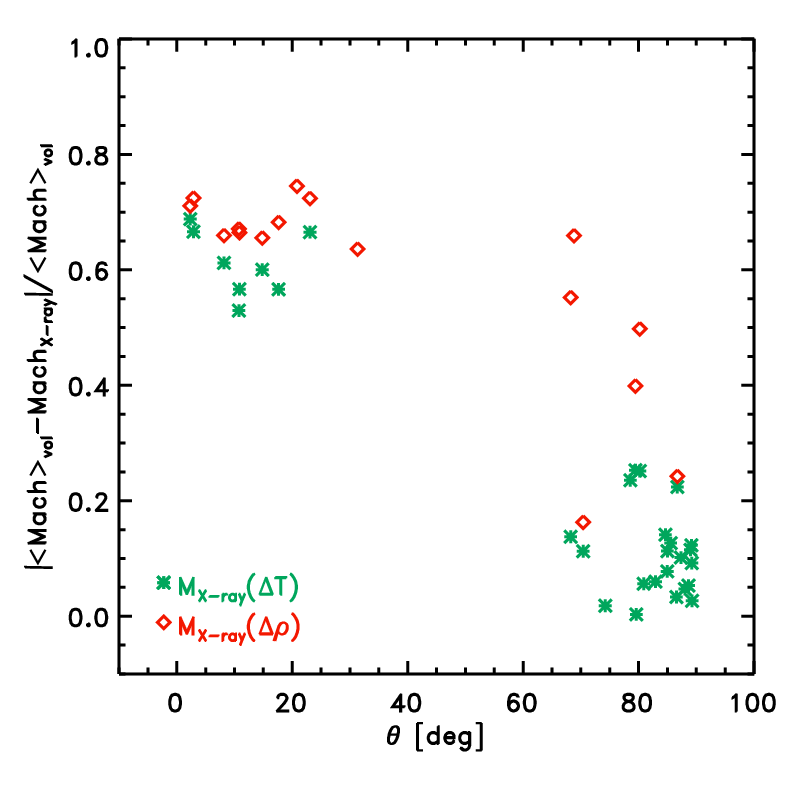}
 \caption{3D analysis, left: X-ray weighted Mach number plotted against the X-ray Mach number. Right: Relative difference between the volume weighted average Mach number and the X-ray Mach numbers plotted against the viewing angle. In both panels, the Mach number has been estimated using the temperature jump (green asterisks) and the density jump (red diamonds).}
 \label{fig::obs_vs_dist_2}
\end{figure*}

 In Fig. \ref{fig::obs_vs_dist}, we compare the radio Mach numbers, that were computed from the integrated radio spectral index, to the radio weighted and volume weighted average Mach numbers. The average of the radio weighted distribution and the radio Mach number are very close to equality, as it should be according to the simulation set up. Consequently, the radio Mach number is always larger than the volume weighted average and the difference between the two increases for brighter relics. The differences are of the same size as one standard deviation of the Mach number distributions. For less powerful relics, i.e. $P_{1.4 \ \GHz} < 10^{30} \erg \, \sek^{-1} \, \Hz^{-1}$, the difference is even below one standard deviation. 
 
 On the other hand, the X-ray Mach numbers, computed from the measured temperature or density jump, are mostly smaller than the corresponding X-ray weighted average (see Fig. \ref{fig::obs_vs_dist_2}).  Furthermore, the X-ray Mach numbers are found to be mostly smaller than the volume weighted average. We did not find a correlation between the radio power and the difference between the X-ray Mach numbers and the volume weighted average. However, we found a correlation between the difference of the two and the orientation of the relic. The last panel in Fig. \ref{fig::obs_vs_dist_2} displays, how the relative difference between the volume weighted average and the X-ray Mach number depends on the viewing angle. If the X-ray Mach number is computed from the temperature jump, the relative difference decreases for relics that are seen close to edge-on. On the other hand, this trend is not observed for the density jump based X-ray Mach numbers. Consequently for relics seen edge-on, the temperature jump based X-ray Mach number appears to be a good match for the average value of the underlying Mach number distribution. Analyses by \citep{Dawson_2015} and \citep{Golovich_2017}, showed that most relics are seen edge-on, as the merger axis are near the plan of the sky. Hence, for the majority of observed relics the X-ray Mach numbers should mirror the average of the underlying distribution.

\section{Discussion}\label{sec::summary}

We have used numerical simulations to investigate the long-standing issue of the discrepancy between the Mach numbers in cluster shocks estimated with X-ray observations and and their corresponding radio Mach number, which is inferred from the integrated radio spectrum  \citep[e.g.][]{2011MmSAI..82..495M,2013PASJ...65...16A,2016MNRAS.461.1302E,2016ApJ...818..204V,2017A&A...600A.100A,Hoang2017sausage}. In particular, we assumed a diffusive shock acceleration model to power radio relics in the simulations, and tested whether the discrepancy follows from the fact that the different Mach number measurements probe different parts of the underlying complex Mach number distribution.

\subsection{Synthetic shock fronts}

As a first step, we studied the Mach number discrepancy in a controlled environment, see Section \ref{sec::1D}. To first order, the radio power and, hence, the radio spectrum depend solely on the Mach number of the shock. If the Mach number at shocks is not single-valued, but there is a distribution of shock strengths, the measured radio Mach number is always biased towards the high value tail of the distribution. This finding is robust for all plausible variations of the (still very uncertain) shock acceleration efficiency of electrons as a function of the shock strength. We also showed that the X-ray Mach number is always biased towards the mean of the underlying Mach number distribution. Consequently, it is plausible that there is an intrinsic discrepancy between the observed radio and X-ray Mach numbers.

\subsection{Cosmological simulations}

We expanded our analysis to a more realistic sample of 3D radio relics that were simulated with the cosmological code \enzo, see Section \ref{sec::3D}. For each relic of our sample, we computed the Mach number based on the temperature/density jump observed in X-rays and from the integrated radio spectral index. If the radio Mach number and at least one X-ray Mach number were available, we computed the corresponding Mach number discrepancy. Our findings are summarized as follows

\begin{enumerate}

    \item \textit{The X-ray Mach number measured in the simulations are in line with typically observed numbers}. In a few cases, our simulated X-ray Mach numbers obtained from the temperature jump are below the corresponding density jump based measurement. This behaviour is not observed in the observational sample. Moreover, we found that \textit{the X-ray Mach number is very sensitive to the orientation of the relic to the observer}. Already, changes of a few degrees can alter the measured X-ray Mach numbers. See Sections \ref{ssec::xray_mach_number}, \ref{sssec::xray_ratio} and Fig. \ref{fig::discrepancy}. \\
    
    \item \textit{The radio Mach numbers in the simulations are found in the upper part of the corresponding observed distribution of shock strength.} Our simulated relics sample does not include any steep spectrum relics and, hence, all estimated radio Mach numbers are rather strong. The observed relic sample also includes relics with steeper spectra and, hence, smaller radio Mach numbers. See Section \ref{ssec::radio_mach}, Fig. \ref{fig::comparison} and \ref{fig::discrepancy}.
    
\end{enumerate}

The simulated sample might not contain any steep spectrum sources for several reasons. The simulations might fail to reproduce such sources, because we did not model the energy gains and losses of the accelerated electrons, e.g. using a Fokker-Planck solver \citep[e.g.][]{vazza2021radiogal}. Furthermore when compiling the simulated sample, we only included bright radio relics. In order to produce such bright relics in simulations, strong shock waves are needed. This bias could be overcome, if one includes the effect of re-acceleration in the simulations. An other cause could be that we did not mimic any of the observational challenges, such as detection limits or exposure time, when computing the radio spectra. On the other hand, it is also possible that the observations are underestimating the spectral indices. For the majority of relics, the observed spectrum was measured at three frequencies. If one of these measurements is shallow, improperly deconvolved, or performed without any matching uv-coverages, the observed spectrum might appear steeper. Recent observations by \citet{loi2020sausspec}, \citet{rajpurohit2020toothspec} and \citet{rajpurohit2021A2744} highlighted the importance of enough frequency coverage, when computing integrated radio spectra. In addition, proper subtraction of unrelated sources is very important which may also steepen the spectrum. However, to fully understand the differences between the observed and simulated radio Mach numbers, more and deeper observations as well as more sophisticated simulations, that carefully follow the spectral evolution of the cosmic-ray electrons, are required.

\begin{enumerate}
  \setcounter{enumi}{3}
    \item \textit{In the simulation, we find a Mach number discrepancy that is in good agreement with the one found in observations.} There is little evidence that the Mach number discrepancy could increase for brighter relics and more X-ray luminous clusters. Except for two cases, the radio Mach numbers in the simulations are always larger than the corresponding X-ray Mach numbers. See Section \ref{sssec::discrenpancy}, Fig. \ref{fig::discrepancy} and \ref{fig::xray_vs_diff_2}. \\
    
    \item \textit{In the simulations, the relative discrepancy between the X-ray and radio Mach numbers depends on the orientation of the relic.} The discrepancy is smaller for relics that are seen edge-on and it increases for relics seen face-on. These findings highlight the sensitivity of the X-ray measurement to the projection of the relic: If a relic is seen at an oblique angle, the observed temperature jump might appear weaker and, hence, the Mach number can be underestimated. Furthermore, variations in the gas density such as clumps can affect the X-ray measurement. Finally, relics are curved and extended structures. Thus, the direction of the shock normal varies across a relic's surface. A projected map can not completely characterise such a shock front. See Section \ref{sssec::discrenpancy} - \ref{sssec::xray_ratio}, Fig. \ref{fig::discrepancy} and  \ref{fig::xray_vs_diff_2}. \\
    
    \item In agreement with works by \citet{2011JApA...32..509H,2013ApJ...765...21S,2015ApJ...812...49H} and \citet{Roh2019}, \textit{the simulated relics trace Mach number distributions that peak at low Mach numbers and show a high Mach number tail.} Computing the distributions using various weights showed that the radio weighted and X-ray weighted distributions are different. The radio weighted distribution is dominated by the few strong Mach numbers that reside in the high value tail of the distribution. On the other hand, the X-ray weighted distribution peaks at low values. This difference has also been found by \citet{Dominguez_2020_relicsI}, who argued that the radio emission can only probe parts of the shock due to its patchy nature and magnetic fluctuations in the downstream, that decrease more slowly than the fluctuations of density, temperature and velocity. (Section \ref{ssec::distributions}, Fig. \ref{fig::main_results} - \ref{fig::obs_vs_dist_2})
    
\end{enumerate}

 \textit{Our findings suggests that X-ray and radio Mach numbers differ intrinsically as they trace different parts of the underlying Mach number distribution.} If perfectly measured, the X-ray Mach number reflects the bulk of values of the Mach number distribution, which are related to the total pressure jump produced by the shock. On the other hand, the radio Mach number is always larger than the average of the distribution and it resides in the distribution's high value tail. However, the difference between the radio Mach number and the volume weighted average Mach number is of the order of one standard deviation of the underlying Mach number distribution. Consequently, the radio Mach number might reflect the width of the underlying Mach number distribution. 
 
 This brings up the idea that one can estimate the actual underlying Mach number distribution using the X-ray and radio Mach number, if both are perfectly measured. As we have shown, the measured X-ray Mach number is sensitive to projection effects and the applied method. Consequently, it is difficult to make a conclusion about the underlying Mach number distribution based on the X-ray observations.
 
 High resolution observations of radio relics show that the radio spectral index varies across the shock front on scales of a few $10 \ \kpc$ \citep[e.g.][]{rajpurohit2020toothbrush,digennaro2018saus}. These variations might provide information about the Mach number distribution across the shock. Methods that spatially resolve radio relics, i.e. such as colour-colour plots or spectral tomography, should be more suitable for such a study. Our first explorative study using cosmological simulations showed that the local spectral indices are also biased towards the stronger shocks along the line-of-sight. However, directly connecting the local spectral index to the underlying Mach number is not so trivial, as Eq. \ref{eq::ainj_mach} is not applicable. Thus, this will be the subject of future work. \\
 
 Generally speaking, our analysis suggests that the measurement of the radio Mach number is more robust, as it is less prone to projection effects. However, provided that a relic is observed close to edge-on, the Mach number inferred from X-ray jumps is the quantity that best constraints the average shock strength and that it can be best connected with the large-scale jump of thermodynamical quantities. To determine the orientation of an observed relic, one has to rely on additional measurements, e.g. the spectral index gradient or the degree of polarisation \citep[e.g.][]{ensslin1998}. \textit{In this work, we have introduced a new observational proxy to determine a relic's orientation: the X-ray luminosity ratio, see Section \ref{sssec::xray_ratio}.}
 
 \subsubsection{Numerical limitations}

 Our model of relativistic electrons includes some crude assumptions, that will be improved in the future. First, the model assumes that the properties of the shock front do not change within the electron cooling time. Consequently, both the downstream temperature and the magnetic field strength at the shock determine the downstream radio profile. This is a crude assumption as both quantities might affect the profiles shape. However in most cases, the downstream width at frequencies above $1.0 \ \GHz$ is below the simulation's physical resolution and the effective resolution of the Dedner-cleaning procedure. Therefore for frequencies above $1.0 \ \GHz$, constant shock properties are a valid assumption. 
 
 Second, our model does not include the injection of magnetic fields and relativistic electrons from active galactic nuclei, which could affect the relics' shapes and sizes \citep{2017MNRAS.470..240N}. Finally, in reality, the exact shapes of the electron spectrum and the radio profile depend on the exact values of $E_{\min}$ and $E_{\max}$, which are determined by the microphysics of the plasma as well as by the local shock dynamics, e.g. the magnetic field obliquity. However, at present their values can only be  loosely constrained by kinetic simulations or semi-analytical models \citep[e.g.][]{Ha2018protons,2019ApJ...876...79K,2020JKAS...53...59K}. Despite their advances, even kinetic simulations are not yet able to provide the exact values. An important goal of future studies will be to further improve these.

\section{Conclusions}\label{sec::conclusion}

 In this work, we confirmed that the radio Mach numbers measured for radio relics are on average higher than the corresponding X-ray Mach numbers. To this end, we have compiled an up-to date sample of observed relics for which both radio and X-ray Mach number estimates are present. We demonstrated that X-ray and radio mock observations of shock fronts in merging clusters simulated with high resolution in a cosmological environment show a similar discrepancy. With a detailed analysis, we showed that the discrepancy is attributed to three effects:

\begin{enumerate}

 \item The high Mach number regions of the shock front are significantly more radio luminous. \\

 \item X-ray surface brightness analysis may suffer from projection effects, while radio measurements are basically unaffected by projection. \\

 \item Weaker Mach numbers tend to be found in X-ray brighter regions.
 
\end{enumerate}

 \citet{Dominguez_2020_relicsI} showed that the underlying Mach number distribution depends on two parameters: the initial strength of the shock wave and the size and strength of fluctuations in the upstream gas. Their results suggest, that the peak of the Mach number distribution depends on the initial shock strength (compare with figure B2 in \citet{Dominguez_2020_relicsI}). The width of the distribution is characterised by the amount of ICM fluctuations.
 
 In combination with the results by \citet{Dominguez_2020_relicsI}, we can now propose the following scenario, which plausibly addresses the issue of the "Mach number discrepancy", reported by observations: shocks that form during mergers in the ICM are characterised by an initial strength. However due to fluctuations in the ICM, the Mach number varies across the shock front, producing a broadened distribution of values. The peak of this distribution is connected with the strength of the input Mach number. 
 During the shock propagation, if a X-ray Mach number can be measured (in the absence of projection effects) it  matches the peak of the Mach number distribution. The width of the distribution depends on the ICM fluctuations. The shocks in the high value tail are the one responsible for most of radio emission, and thus they  dominate the radio spectrum. Consequently, the radio Mach number is biased to the high value tail and, therefore, mirrors the width of the Mach number distribution. However, our analysis also showed that, for low power radio relics, the radio Mach numbers can also underestimate the width of the Mach number distribution, see Fig. \ref{fig::obs_vs_dist}. This makes these findings less generalisable and future studies, targeting the connection between the ICM fluctuations, the width of the Mach number distribution and the radio Mach number, are needed.

\section*{acknowledgements}
  We thank our anonymous referee for the helpful comments.  \\
  The authors gratefully acknowledge the Gauss Centre for Supercomputing e.V. (www.gauss-centre.eu) for supporting this project by providing computing time through the John von Neumann Institute for Computing (NIC) on the GCS Supercomputer JUWELS at J\"ulich Supercomputing Centre (JSC), under project no. hhh44. \\
  D.W. is funded by the Deutsche Forschungsgemeinschaft (DFG, German Research Foundation) - 441694982. 
  S.E. acknowledges financial contribution from the contracts ASI-INAF Athena 2019-27-HH.0, ``Attivit\`a di Studio per la comunit\`a scientifica di Astrofisica delle Alte Energie e Fisica Astroparticellare'' (Accordo Attuativo ASI-INAF n. 2017-14-H.0), INAF mainstream project 1.05.01.86.10, and funding from the European Union’s Horizon 2020 Programme under the AHEAD2020 project (grant agreement n. 871158).
  K.R and F.V. acknowledges financial support from the European Union's Horizon 2020 program under the ERC Starting Grant "MAGCOW", no. 714196. 
  M.H. acknowledges support by the BMBF Verbundforschung under the grant 05A20STA.
  P.D.F. was supported by the National Research Foundation (NRF) of Korea through grants 2016R1A5A1013277 and 2020R1A2C2102800. \\
  This research made use of the radio astronomical database Galaxy\nolinebreak Clusters.com, maintained by the Observatory of Hamburg. Finally, we acknowledge fruitful discussions with M. Br\"uggen.
 \section*{Data availability}
 The data underlying this article will be shared on reasonable request to the authors, unless being in conflict or breaking the privacy of ongoing related work led by our collaboration members.
 \bibliographystyle{mnras}
 \bibliography{mybib}

\appendix

 \section{Additional synthetic shocks tests}\label{app::conv}
 
 The model applied in Section \ref{sec::1D} can induce a numerical Mach number discrepancy, if the error of the numerical integration of Eq. \ref{eq::dPdVdv} is too large. As a sanity test of the implementation, we computed the radio Mach number for synthetic shock fronts that only consist of a single input Mach number. In these cases, the radio Mach number should resemble the input Mach number. For all different input Mach numbers, the relative difference between the input Mach number and the radio Mach number is $10^{-3}$. Consequently, we have reached numerical convergence and, in the analysis, any deviations between the input Mach numbers and the derived radio Mach numbers are of physical nature.
 
 Following Eq. \ref{eq::ainj_mach}, the spectral index only depends on the underlying Mach number. If the Mach number is constant across a relic's surface, the integrated spectrum is expected to be independent of variations of the magnetic field as well as of the acceleration efficiency. The latter could also vary across the relic's surface, if it does not only depend on the Mach number but also on other quantities, such as the magnetic field obliquity at the shock \citep[e.g.][]{Guo_eta_al_2014_I}. Hence, similar to the approach in Section \ref{sec::1D}, we built synthetic shock fronts. Here, we  used: (1)  a fixed Mach number, $M$, and fixed acceleration efficiency, $\xi$, as well as a distribution of magnetic fields $B$ across the shock front; (2) a fix Mach number, $M$, and fixed magnetic field, $B$, as well as a distribution of acceleration efficiencies, $\xi$. For the input parameters we used:
 
 \begin{itemize}
  \item $B = [0.1-0.9] \ \mu G, \ \mathrm{with} \ \delta B = 0.1 \ \mu G$
  \item[] $B = [1-9] \ \mu G, \ \mathrm{with} \ \delta B = 1 \ \mu G$
  \item $\xi = [0.01,0.05], \ \mathrm{with} \ \delta \xi = 0.01$.
  \end{itemize}
  
 Here, in the input distribution of the magnetic field/acceleration efficiency every values occurs once. In both cases, the input Mach number is constant across the synthetic shock front. Hence, the ``true'' Mach number is always the input Mach number and one would expect that the integrated spectral index matches that input Mach number. For each synthetic shock front, we computed  the relative difference between the input Mach number and the radio Mach number. The relative differences are very small, i.e, below $10^{-2}$. Consequently, as expected, if the Mach number across a relic is constant, but either the magnetic field or the acceleration efficiency fluctuate, the integrated spectral index of the radio emission still re-produces the input Mach number. Hence, in this case no bias is expected in the comparison between the Mach numbers inferred from either X-ray or radio observations.  

\begin{figure*}
  \includegraphics[width = 0.195\textwidth]{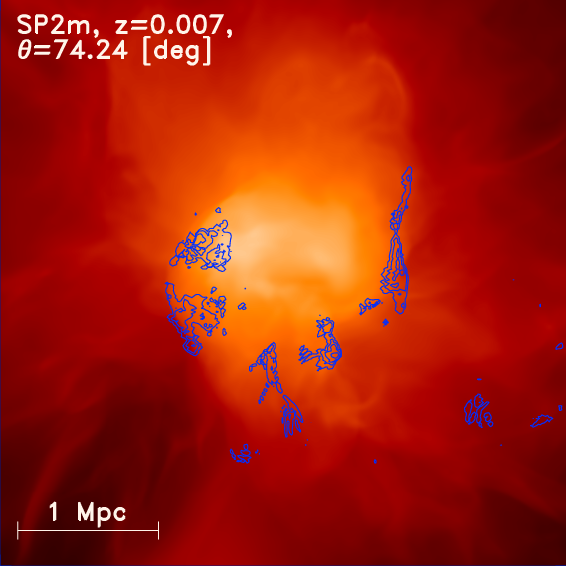} 
  \includegraphics[width = 0.195\textwidth]{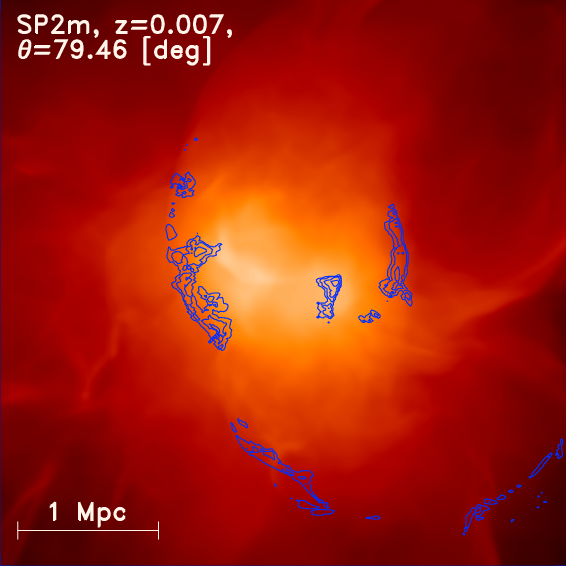}
  \includegraphics[width = 0.195\textwidth]{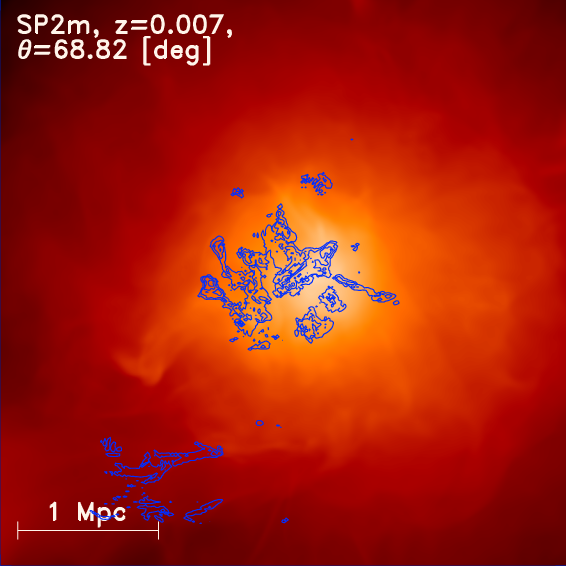}
  \includegraphics[width = 0.195\textwidth]{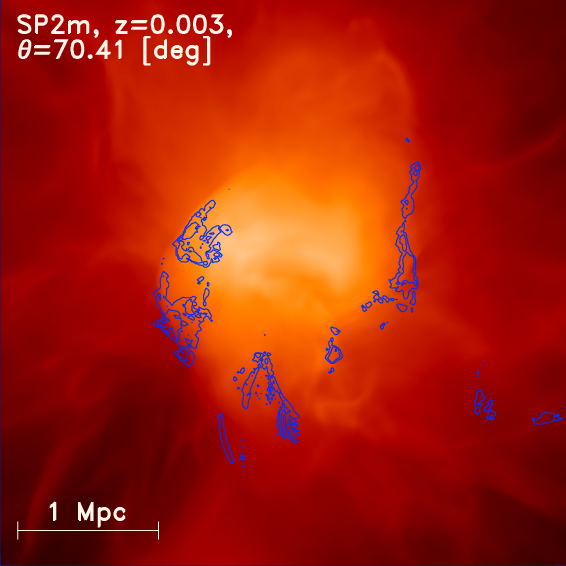} 
  \includegraphics[width = 0.195\textwidth]{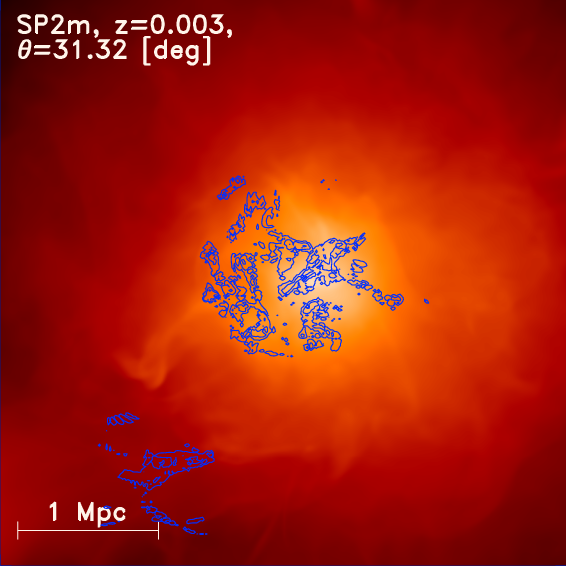} \\
  \includegraphics[width = 0.195\textwidth]{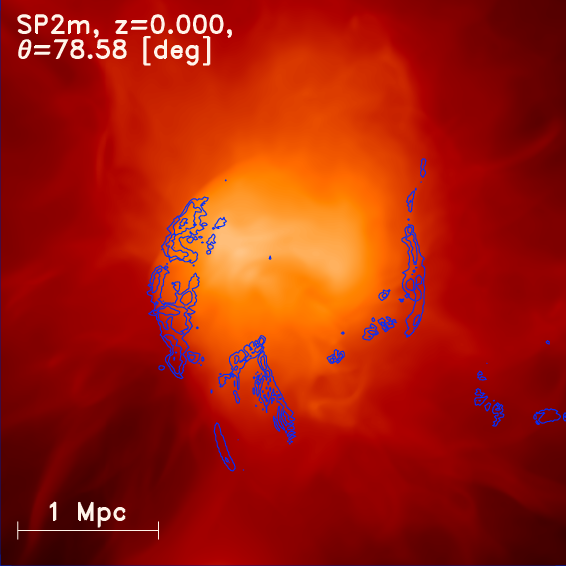} 
  \includegraphics[width = 0.195\textwidth]{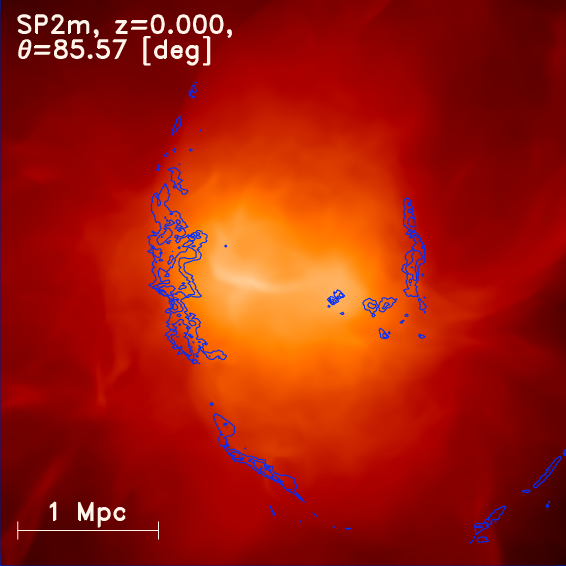} 
  \includegraphics[width = 0.195\textwidth]{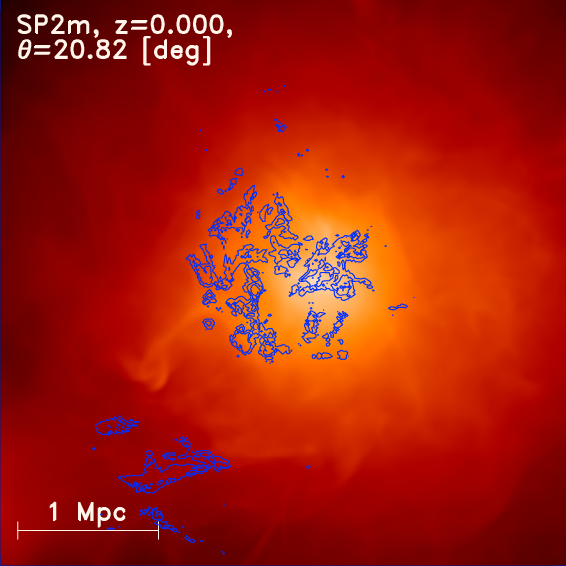}
  \includegraphics[width = 0.195\textwidth]{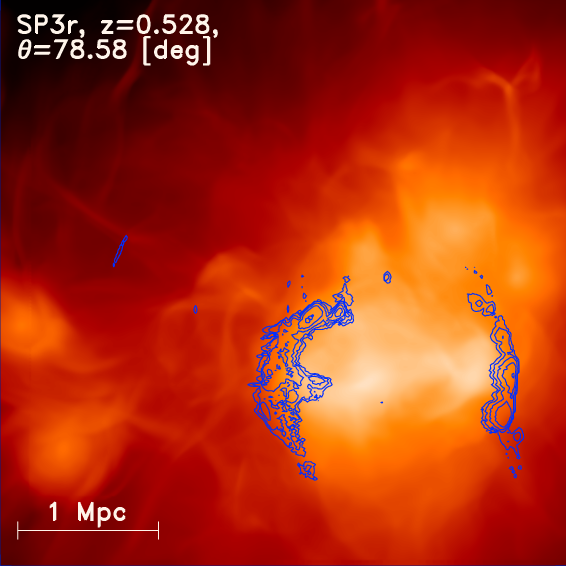}
  \includegraphics[width = 0.195\textwidth]{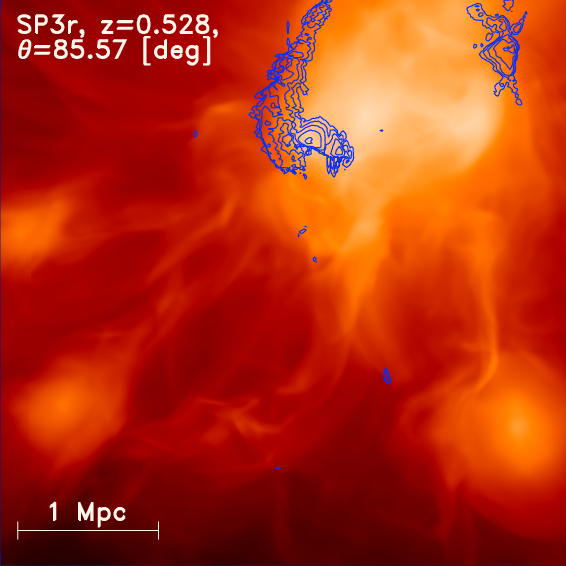} \\
  \includegraphics[width = 0.195\textwidth]{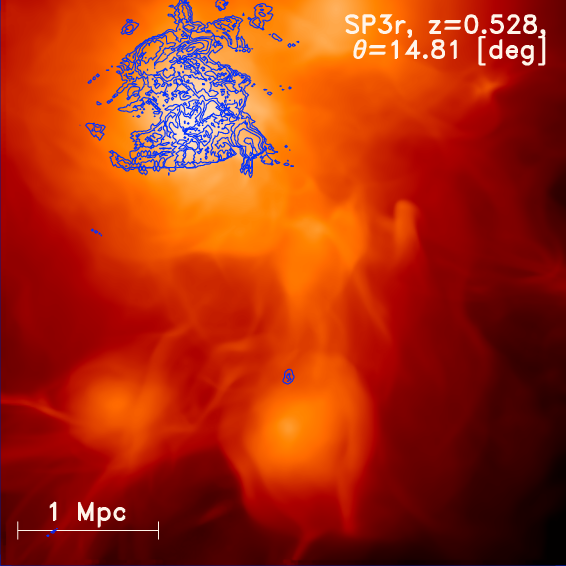} 
  \includegraphics[width = 0.195\textwidth]{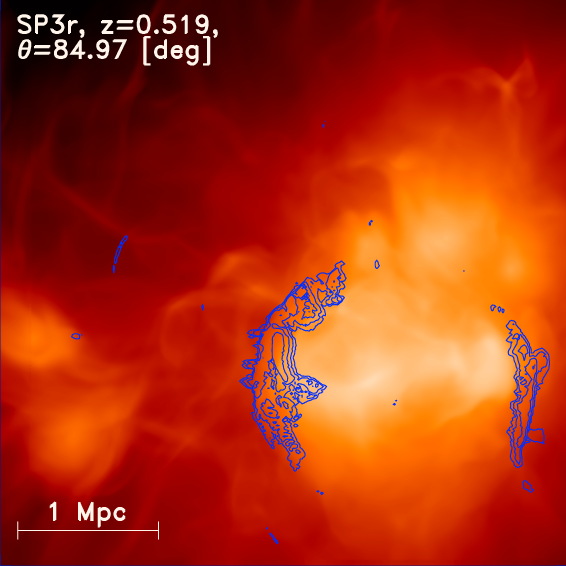}  
  \includegraphics[width = 0.195\textwidth]{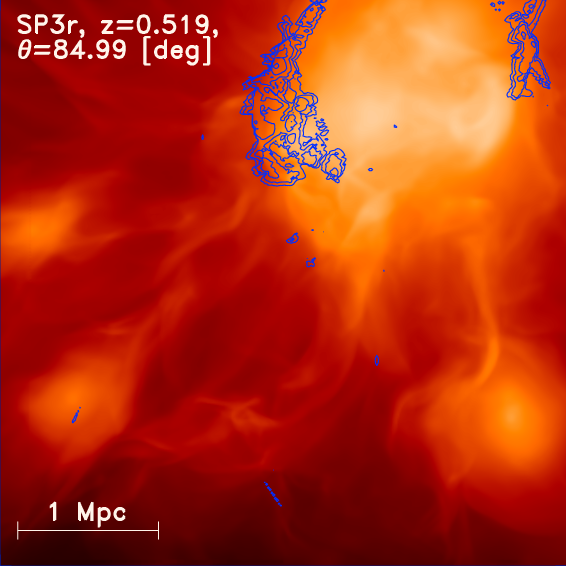}
  \includegraphics[width = 0.195\textwidth]{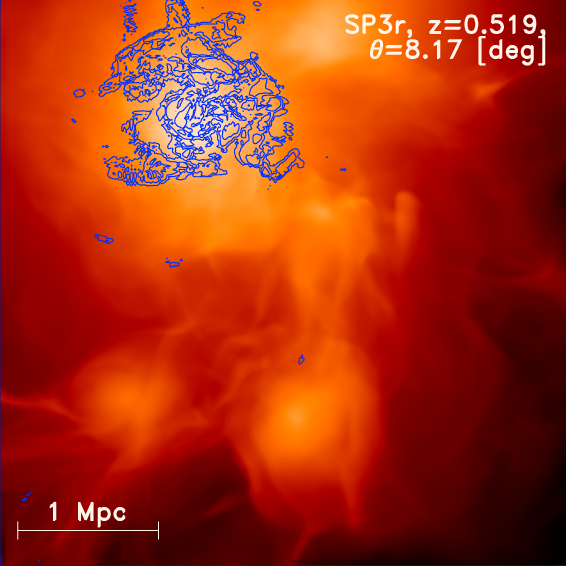}
  \includegraphics[width = 0.195\textwidth]{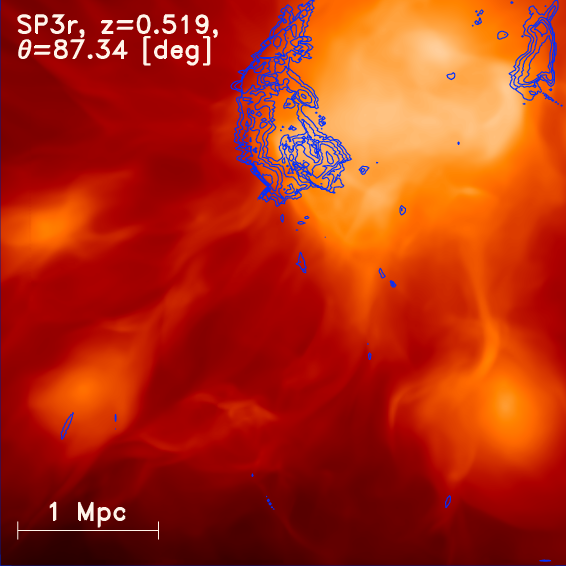} \\
  \includegraphics[width = 0.195\textwidth]{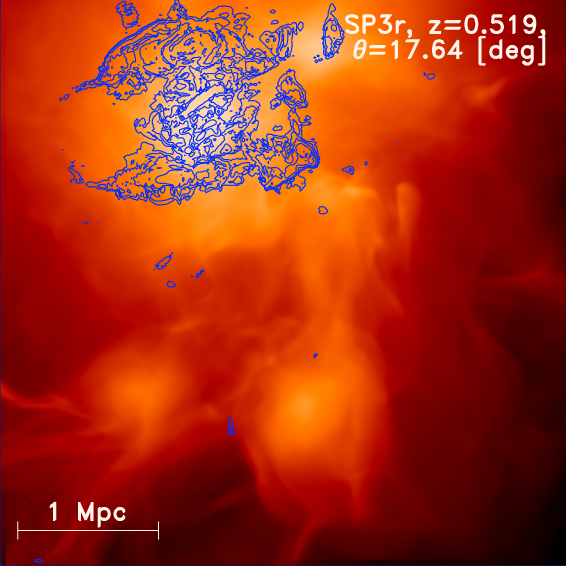} 
  \includegraphics[width = 0.195\textwidth]{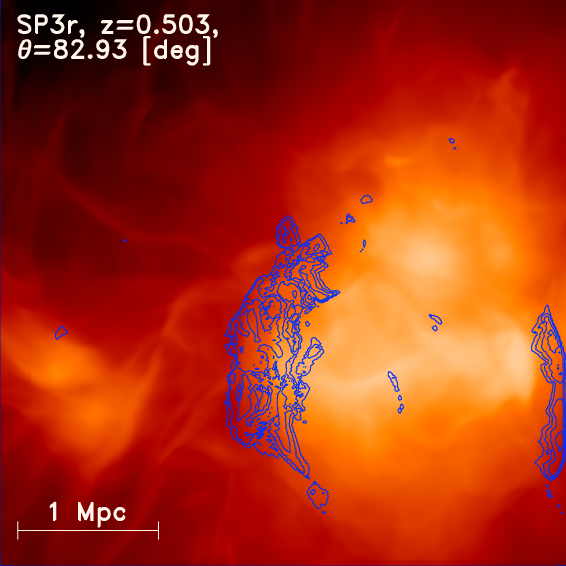} 
  \includegraphics[width = 0.195\textwidth]{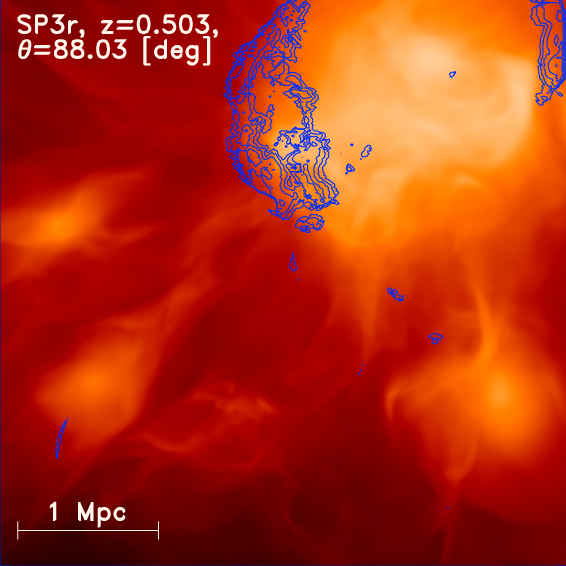}
  \includegraphics[width = 0.195\textwidth]{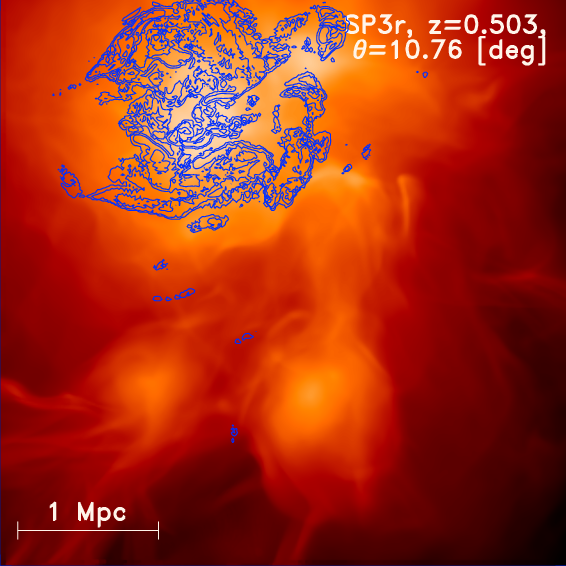} 
  \includegraphics[width = 0.195\textwidth]{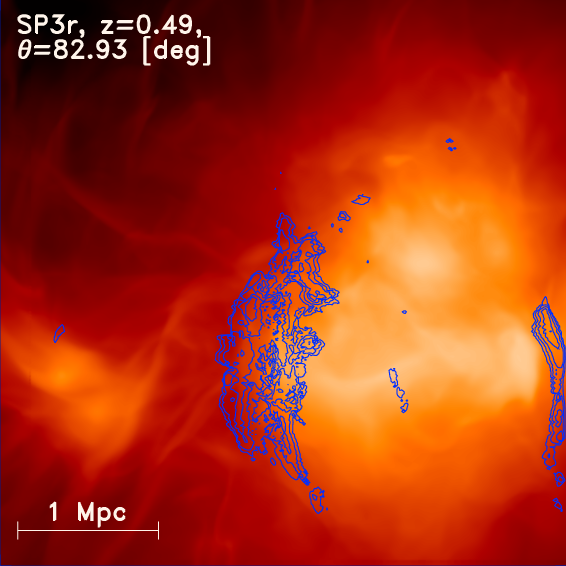} \\
  \includegraphics[width = 0.195\textwidth]{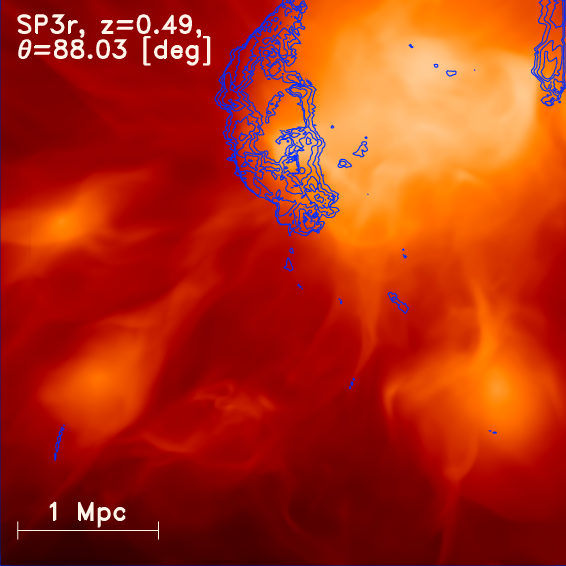} 
  \includegraphics[width = 0.195\textwidth]{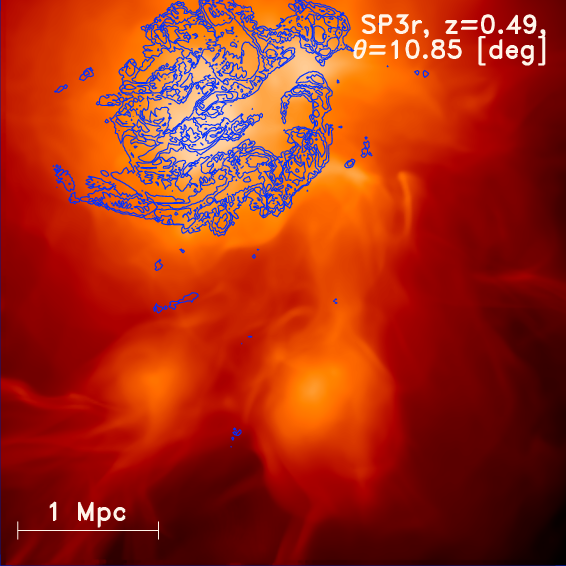}
  \includegraphics[width = 0.195\textwidth]{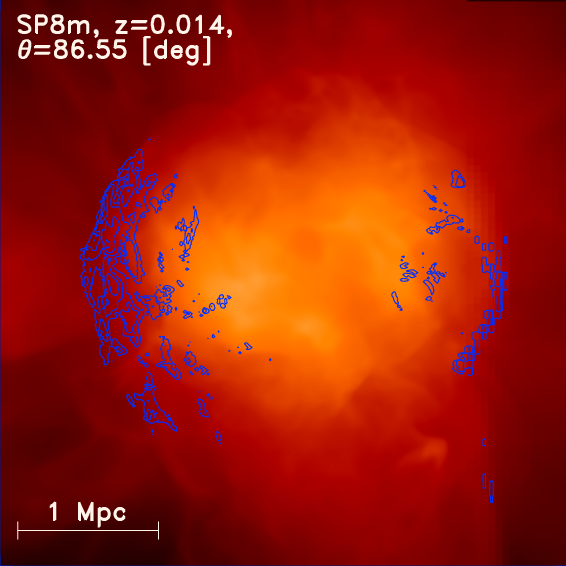} 
  \includegraphics[width = 0.195\textwidth]{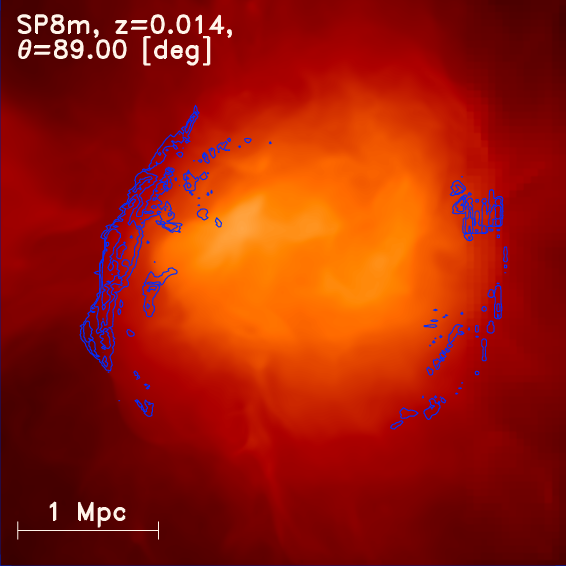}
  \includegraphics[width = 0.195\textwidth]{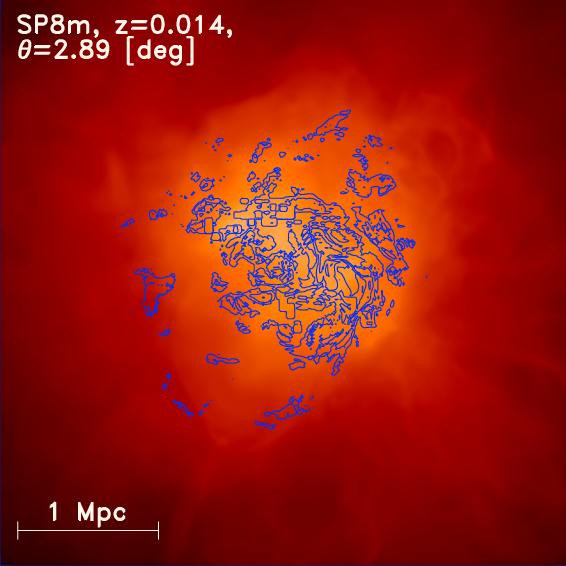} \\
  \includegraphics[width = 0.195\textwidth]{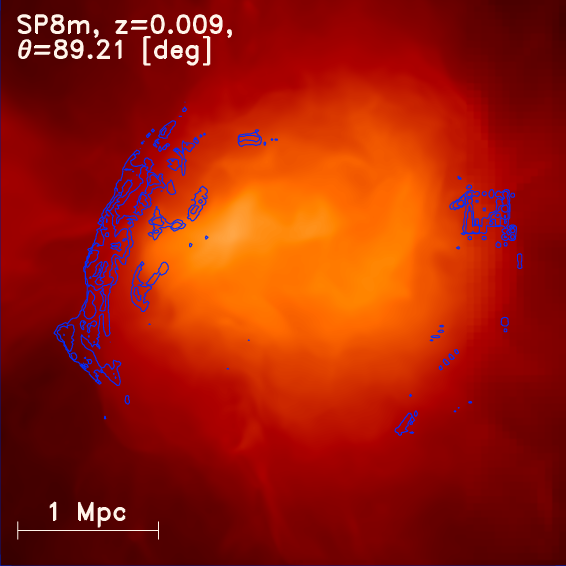}
  \includegraphics[width = 0.195\textwidth]{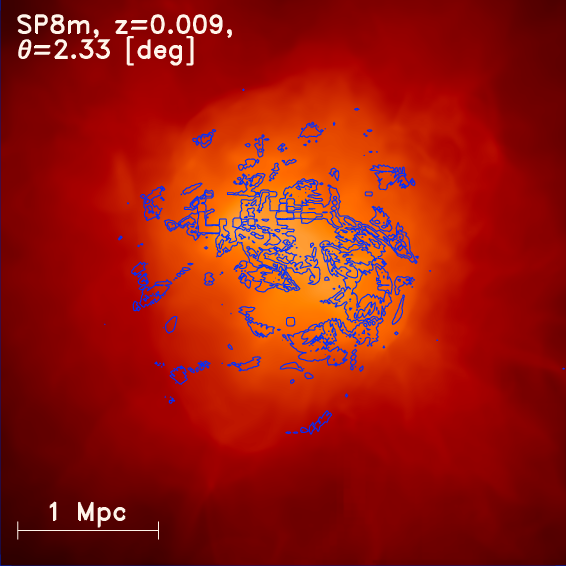}
  \includegraphics[width = 0.195\textwidth]{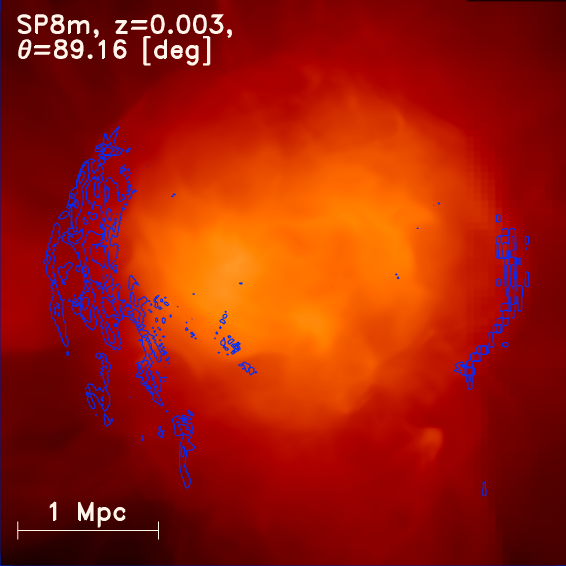} 
  \includegraphics[width = 0.195\textwidth]{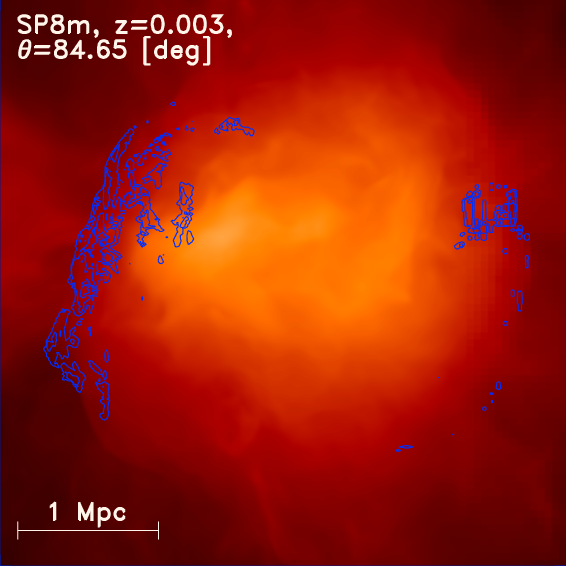}
  \includegraphics[width = 0.195\textwidth]{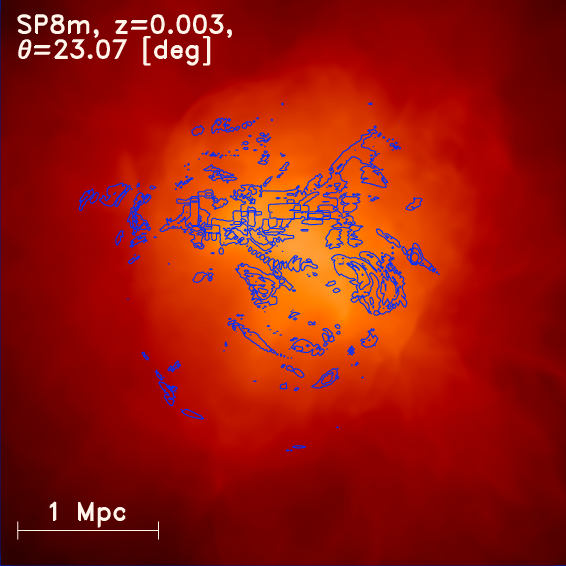}
  \caption{3D analysis: The radio contours at $1.4 \ \GHz$ on top of the X-ray luminosity of our simulated cluster sample, excluding the relic shown in Fig. \protect{\ref{fig::cluster_sample}}. In the maps that show double radio relics, we always analysed the relic on the left side.}
  \label{fig::cluster_sample_all}  
\end{figure*}

\section{The cluster sample}\label{app::clusters}

Excluding the relics shown in Fig. \ref{fig::cluster_sample}, we plot the radio contours of all relics on top of the X-ray emission map of the host cluster in Fig. \ref{fig::cluster_sample_all}.

\begin{figure}
    \centering
    \includegraphics[width = 0.5\textwidth]{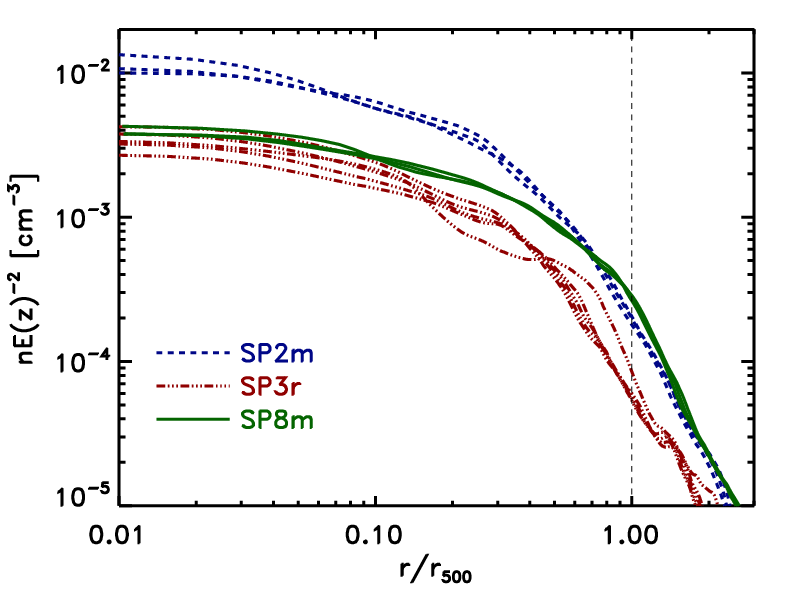} \\
    \includegraphics[width = 0.5\textwidth]{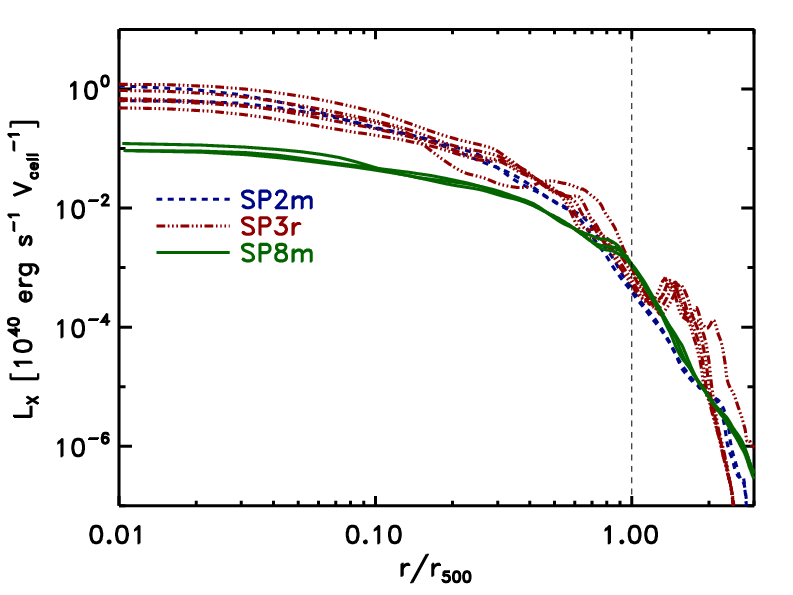}
    \caption{3D analysis: re-scaled density profiles of our simulated sample (top) and X-ray luminosity profile bottom (bottom). The vertical dashed line marks the $r_{500}$.} 
    \label{fig::r500_profiles}
\end{figure}

We compared our cluster sample with the X-COP sample of cluster observed in X-ray, analysed by \citep{Ghirardini:2019}, which contains 12 clusters with masses in the range of $M_{200} \approx 6-15 \cdot 10^{15} \ \Msun$. Consequently, our simulated sample is at the lower mass range of the observed sample. Following \citep{Ghirardini:2019}, we computed the density profile inside the radius $r_{500}$. We re-scaled the density profile by the function $E(z) = \Omega_{\mathrm{M}} (1+z)^3 + \Omega_{\Lambda}$. In Fig. \ref{fig::r500_profiles}, we plot the re-scaled profiles. The density profiles of our simulated sample are less cuspy than the observed sample, which is expected in the absence of radiative gas cooling in our cluster cores. As a consequence, the  X-ray surface brightness profiles in the simulation are flatter around the X-ray peak, compared to observed  real clusters (see Fig. \ref{fig::r500_profiles}).

\section{Acceleration efficiencies}\label{sssec::acceleff}
\begin{figure}
  \includegraphics[width = 0.49\textwidth]{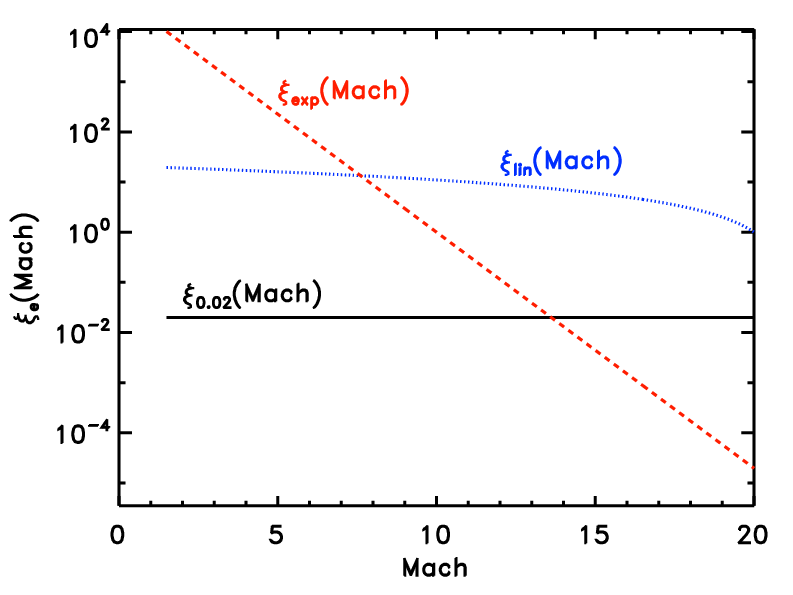}
  \caption{3D analysis: Different functions of $\xi_e(M)$. Acceleration from the thermal pool is given by the constant function $\xi_{0.02}(M)$ (black, solid). The other two functions, $\xi_{\mathrm{lin}}(M)$ (blue, dotted) and $\xi_{\mathrm{exp}}(M)$ (red, dashed), give more power to the low Mach number shocks.}
  \label{fig::efficiencies}
 \end{figure}

 Our findings suggest that the radio Mach number, i.e. that derived from the integrated spectral index, is biased towards the shocks that process most of the energy and, hence, produce the bulk of radio emission. The amount of produced radio emission depends on the assumed acceleration efficiency, i.e. $\xi_e$ in Eq. \ref{eq::cspec}, and the integral in Eq. \ref{eq::dPdVdv}. As discussed above, the integral in Eq. \ref{eq::dPdVdv} rapidly decreases for small Mach numbers. Therefore in our modelling, most of the radio emission is produced by the regions in the shock with the highest Mach number. However, if weaker shocks produce a larger fraction of the observed radio power, the shape of the radio spectrum might change. For example, weaker shocks can process more energy and produce more radio emission if they re-accelerate populations of mildly relativistic fossil electrons \citep[e.g.][]{2013MNRAS.435.1061P}. To incorporate re-acceleration in the computation, one needs to carefully follow the energy gains and losses of the accelerated electrons, e.g. using a Fokker-Planck solver \citep[e.g.][]{vazza2021radiogal}, which is beyond the scope of this work. In our model, we can produce more radio power with weak shocks by modifying $\xi_e(M)$. We used two additional forms for $\xi_e(M)$ in Eq. \ref{eq::dPdVdv}:
 \begin{align}
  \xi_{\mathrm{lin}}(M) &=  -M + 21\\ 
  \xi_{\mathrm{exp}}(M) &= 50802  \exp\left(-\frac{M}{0.922876}\right) .
 \end{align}
 Strictly speaking, these two forms of $\xi_e(M)$ are nonphysical as they suggest efficiencies that are larger than $100 \ \%$. However, if we can show that our results remain the same for such large (and nonphysical) variations in the acceleration efficiency, they will also not change for smaller variations in the acceleration efficiency. In Fig. \ref{fig::efficiencies}, we plot the shape of the different efficiency functions.
 
  The $\xi_{\mathrm{lin}}(M)$-scenario gives a little more power to the weak shocks, i.e. $\xi_{\mathrm{lin}}(1.5)/\xi_{\mathrm{lin}}(20) = 19.5$. The corresponding boost in the total radio power at $1.4 \ \GHz$ is by a factor of $\sim 16$. As discussed in Section \ref{sec::equations}, the integral in Eq. \ref{eq::dPdVdv} decreases rapidly with Mach number and a factor of $\sim 20$ is not enough to balance this decrease. Hence, the function $\xi_{\mathrm{exp}}$ has been chosen, such that the energies processed by a $M = 1.5$ and a $M = 20$ shock are similar. As a consequence, the radio power at $1.4 \ \GHz$ is increased by a factor of $\sim 500$. Using these different efficiencies, we re-computed the radio spectrum for our representative relic. We find that the integrated spectral index changes only slightly. The flatter spectrum is found for $\xi_{\mathrm{lin}}$ with $\alpha_{\mathrm{lin}} = 1.10$ and the steeper for $\xi_{\mathrm{exp}}$ with $\alpha_{\mathrm{exp}} = 1.13$. The associated Mach numbers are $M_{\mathrm{lin}} = 4.58$ and $M_{\exp} = 4.05$, respectively. Both values are still well above the observed X-ray Mach number, which is $M_{\mathrm{X-ray}}(\Delta T) = 2.73$. Consequently, we can conclude that the Mach number discrepancy is largely independent of the underlying efficiencies. 
  Only in the case in which the acceleration efficiency for low Mach numbers is increased even further, the Mach number discrepancy will be alleviated.  However, this seems implausible based on the basic expectations by DSA (also in the case of particle re-acceleration) and it would also yield a corresponding radio power largely exceeding observational constraints. 

 \end{document}